# PHASE SEPARATION, MAGNETISM AND SUPERCONDUCTIVITY IN RUTHENO-CUPRATES

B. Lorenz[1], Y. Y. Xue[1], and C. W. Chu[1,2,3]
[1] Texas Center for Superconductivity, University of Houston, Houston, Texas 77204-5002, USA
[2] Lawrence Berkeley National Laboratory, 1 Cyclotron Road, Berkeley, CA 94720, USA
[3] Hong Kong University of Science and Technology, Hong Kong, China

## 1. INTRODUCTION

The discovery of superconductivity coexisting with weak ferromagnetism in a class of ruthenium-based high-$T_C$ compounds has initiated intensive experimental and theoretical investigations in recent years. Bauernfeind et al. [1] reported the first synthesis of $RuSr_2LnCu_2O_8$ (Ru-1212) and $RuSr_2(Ln_{1+x}Ce_{1-x})Cu_2O_{10}$ (Ru-1222), with Ln=Sm, Eu, and Gd, and the appearance of superconductivity below 45 K in the latter compound for $0.3 \leq x \leq 0.5$. A drop of resistivity, a possible indication of a superconducting transition, was also observed in Ru-1212, but no shielding signal could be detected by the ac susceptibility measurements [1]. The superconductivity in Ru-1222 was independently confirmed by Ono [2] who attributed a strong magnetic signal below 100 K in $RuSr_2Sm_{1.2}Ce_{0.8}Cu_2O_{10}$ to secondary magnetic phases. The coexistence of weak ferromagnetism and superconductivity was first proposed by Felner et al. in $RuSr_2R_{1.4}Ce_{0.6}Cu_2O_{10}$ [3]. It is of particular interest that in these compounds the magnetic transition temperature, $T_m$, is far higher than the superconducting $T_c$ which makes this a unique class of materials as compared to some intermetallic compounds (e.g. $RRh_4B_4$, $RMo_6S_8$, $RNi_2B_2C$) known as magnetic superconductors with $T_m < T_c$. Whereas the signature of superconductivity in Ru-1222 appears to be clear the situation in Ru-1212 is more complicated. The published data for this material spread from samples without any diamagnetic signal in the susceptibility [4], through samples showing zero-resistance and zero-field cooled (ZFC) diamagnetic signal but not field cooled (FC) Meissner signal in the magnetic susceptibility [5-8], and eventually samples with a large FC diamagnetic signal [9]. Although most of the Ru-1212 materials synthesized in different laboratories are chemically and structurally comparable it was shown that the superconducting properties are extremely sensitive to tiny details of sample synthesis and annealing procedures [1, 8]. This may explain some of the contradictory results reported in the literature.

Despite extensive research the magnetic as well as the superconducting orders are far from being understood. In particular, the coexistence of both order parameters and how they accommodate each other is still a matter of controversial discussions. In conventional s-wave superconductors the Cooper pairs are easily broken by local magnetic moments and superconductivity is completely suppressed by the presence of as little as 1 % magnetic impurities [10]. This pair breaking mechanism may be less effective if the crystallographic site of the magnetic ion is well isolated from the conduction path, weakening the interaction between the local magnetic moment and the Cooper pairs. This "spatial" separation of magnetic moments and conduction electrons obviously explains the observed coexistence of antiferromagnetism (AFM) ($T_N<T_C$) and superconductivity in ternary intermetallic compounds [11] and borocarbide systems [12]. The competition between the pair breaking magnetic forces and the superconductivity may even lead to exotic effects like the resistive reentrant behavior observed at zero field in $HoNi_2B_2C$ [12]. In the rutheno-cuprates the Ru ions are also well separated from the $CuO_2$ planes where Cooper pairs are formed. However, the major difference to the aforementioned intermetallic compounds is that a more-or-less homogeneous ferromagnetic (FM) order parameter was suggested by dc magnetic susceptibility measurements over a broad temperature range (and far above $T_c$). This FM long-range order, if homogeneous on a microscopic scale, generates an internal magnetic field at the $CuO_2$ planes that may break the Cooper pairs and destroy superconductivity if the internal field is larger than the critical field. In order to coexist with superconductivity, however, the FM order parameter should be non-uniform in a domain-like or spiral structure [13]. In fact, a spiral FM structure was found in $ErRh_4B_4$ [14] and a domain-like FM structure was observed in $HoMo_6S_8$ [15] and $HoMo_6Se_8$ [16], both coexisting with superconductivity. In these examples the magnetic ordering always develops in the superconducting phase, i.e. $T_m<T_C$. For the superconducting ferromagnet $RuSr_2GdCu_2O_8$ Pickett at al. [17] discussed various physical conditions under which superconductivity and ferromagnetism could coexist. They came to the conclusion that the major limiting mechanism for singlet superconductivity is the electronically mediated exchange field that splits the majority and minority Fermi surfaces and that the superconducting order parameter could develop a spatial variation with non-zero total momentum as described by Fulde-Ferrell-Larkin-Ovchinnikov type theories [18].

Another interesting aspect in the coexistence of FM and superconductivity in type II superconductors is the possible formation of a "spontaneous vortex phase" (SVP) due to the interaction between the wave function of the Cooper pairs and the magnetic field [19-21]. If the internal magnetic field generated by the spontaneous FM order is smaller than the upper critical field ($H_{c2}$) but larger than the lower critical field ($H_{c1}$) of the superconducting state a vortex lattice state can be formed forcing the internal magnetic field to penetrate the superconductor as quantized flux lines. The formation of a SVP in $RuSr_2R_{1.5}Ce_{0.5}Cu_2O_{10}$ (Ru-1222) was proposed by Sonin and Felner [22] and various possible scenarios have been discussed in detail by Chu et al. [23]. Depending on the temperature dependence of the internal field, $4\pi M$ (M = magnetization), and the lower critical field, $H_{c1}$, the SVP may exist below $T_C$ to zero temperature or transform into a Meissner state at a temperature $T_{MS}<T_C$ if $H_{c1}$ increases above $4\pi M$. For Ru-1212 and Ru-1222 the FM transition sets in at far higher temperature than superconductivity ($T_m>T_C$). At $T_C$, according to the SVP model, the FM order parameter should survive and the long-range correlation between the magnetic moments of different RuO planes is established via the flux lattice penetrating the superconducting $CuO_2$ layers.

To better understand the complex interaction between magnetic and superconducting orders it is of particular interest to gather information about both the magnetic order (above and below $T_C$) and the superconducting one. Most recent work to explore the magnetic structure was focused on the Ru-1212 compounds, $RuSr_2(Gd/Eu)Cu_2O_8$. The internal magnetic field, $4\pi M$, was probed by zero-field muon-spin rotation experiments above as well as below $T_C$ [24]. It was found that the FM order sets in at about 133 K in accordance with dc susceptibility measurements and that the FM order parameter was not altered significantly in passing into the superconducting phase. The zero-temperature internal field at the muon site was extrapolated to be about 700 G. However, neutron powder-diffraction experiments have revealed that the magnetic order is predominantly antiferromagnetic with antiparallel Ru moments in all three crystallographic directions (G-type AFM) [25-27]. The estimated magnetic moment, $\approx 1.18$ $\mu_B$, of the Ru ion along the c-axis is compatible with the low spin state of $Ru^{4+}$, although NMR experiments on $RuSr_2YCu_2O_8$ suggest that the Ru ion may be in a mixed valence state with 40 % $Ru^{4+}$ (S=1) and 60 % $Ru^{5+}$ (S=3/2) [28]. The net FM component of the order parameter was estimated to be not larger than 0.1 $\mu_B$ per Ru ion by NPD that is appreciably lower than the local magnetic moment of the Ru detected by NMR [25]. Based on the magnetic scattering experiments a canted arrangement of the AFM Ru moments with a small net FM moment perpendicular to the c-axis has been proposed [27]. Whereas superconductivity in the $CuO_2$ planes may well coexist with the AFM ordered Ru moments the accommodation of a large FM component is still a puzzling issue. For the sister compound, Ru-1222, the exploration of the magnetic structure is far less advanced. Susceptibility measurements show several anomalies (at about 180 K and 90 K in $RuSr_2(Gd/Ce)_2Cu_2O_{10}$) indicating a more complex magnetic ordering in this compound [3]. Superconductivity appears below 45 K with a stronger diamagnetic signal in field cooling experiments than observed in the Ru-1212 system [29].

One of the most disputed questions about superconducting ferromagnets is whether the superconducting state is microscopically homogeneous. This discussion was sparked by the fact that the majority of the Ru-1212 and some of the Ru-1222 samples synthesized by different groups did not show a distinct diamagnetic signal in the field cooled susceptibility, i.e. the Meissner effect typical for a bulk superconductor [3, 7, 8, 24, 30-33]. Evidence for a large Meissner signal in $RuSr_2GdCu_2O_8$ was later reported, but only under very low applied magnetic field and below a temperature $T_{MS}$ much smaller than the onset of superconductivity [9]. The data, therefore, was discussed in terms of a transition from SVP to the Meissner state [23]. Additional evidence for bulk superconductivity was derived from specific heat measurements [34, 35]. Tallon et al. [34] reported a specific heat anomaly at $T_C$ of $RuSr_2GdCu_2O_{10}$ with an unusual magnetic field dependence – the peak of the electronic specific heat at $T_C$ shifted to higher temperature with increasing field. This atypical behavior for a superconducting phase transition was not confirmed by the data of Chen et al. [35]. The thermodynamic transition into the superconducting state, the nature of superconductivity, the details of the magnetic order above and below $T_C$, and the mechanisms of coexistence of superconducting and FM order in the superconducting ferromagnets remain a matter of discussion.

In this work we provide experimental evidence for the occurrence of phase separation in the magnetic phase of the Ru-1212 and Ru-1222 compounds. The predominantly AFM domains are separated by nanoscale FM domains. This explains both, the AFM order observed in neutron scattering experiments and the weak FM signal detected in the dc susceptibility measurements. The magnetic domains are of typical nanometer size and are formed within the grains of the material. The superconductivity then develops in the AFM domains that are coupled across the non-superconducting FM regions by the Josephson effect. The onset of the intra-grain superconductivity is explained as the phase-lock transition of an array of Josephson junctions.

## 2. SYNTHESIS AND STRUCTURE

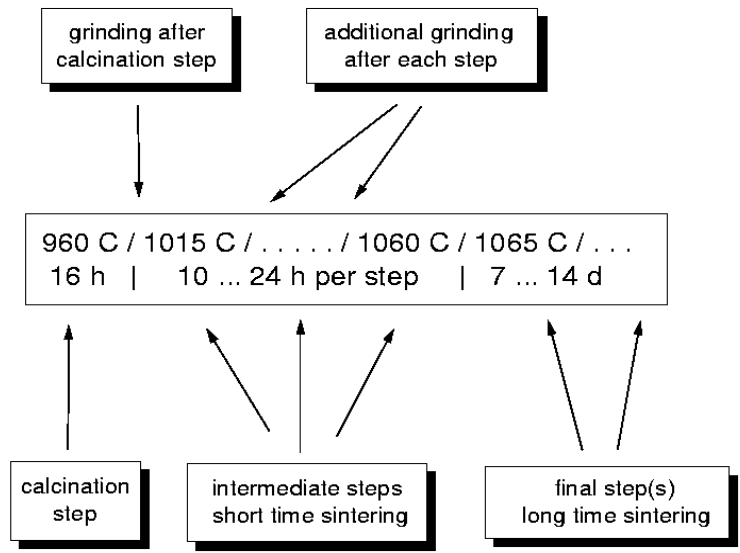

**Fig. 1**: Synthesis and sintering process for $RuSr_2GdCu_2O_8$. The upper line in the central box indicates the temperature and the lower line shows the duration of each step.

The synthesis of Ru-1212 and Ru-1222 compounds proceeds via the solid-state reaction of the proper constituents. Since it was shown that the magnetic and, in particular, the superconducting properties strongly depend on the details of the synthesis and annealing procedures we review the basic steps in this paragraph. As an example, we describe the synthesis procedure of $RuSr_2GdCu_2O_8$. Fig. 1 explains the details of the typical synthesis process as described e.g. in [1, 8]. The starting materials $RuO_2$, $Gd_2O_3$, $SrCO_3$, and CuO were first preheated at 600 – 800 °C for about 12 h. The powder with cation ratio Ru:Sr:Gd:Cu = 1:2:1:2 was then mixed and calcined at 960 °C for 16 h followed by grinding, compacting and two or more sintering steps (each step 10-24 hours) at successively increasing temperature. The final steps include the long-term sintering (7 – 14 days) in oxygen atmosphere. The optimal temperature for this final procedure was estimated to about 1065 °C. After the calcination step the characteristic x-ray reflections for the 1212 phase are already visible in the powder diffraction spectra. Figure 2 shows two examples of x-ray spectra taken after the starting materials were calcined at 960 °C (spectrum A in Fig. 2) and at 1000 °C (spectrum C in Fig. 2). The dots indicate the reflections of the 1212 structure and the crosses show reflections of a secondary phase (identified as $SrRuO_3$). With additional sintering steps the $SrRuO_3$ phase disappears and the spectra B and D in Fig. 2 show the high phase purity of the final product, $RuSr_2GdCu_2O_8$.

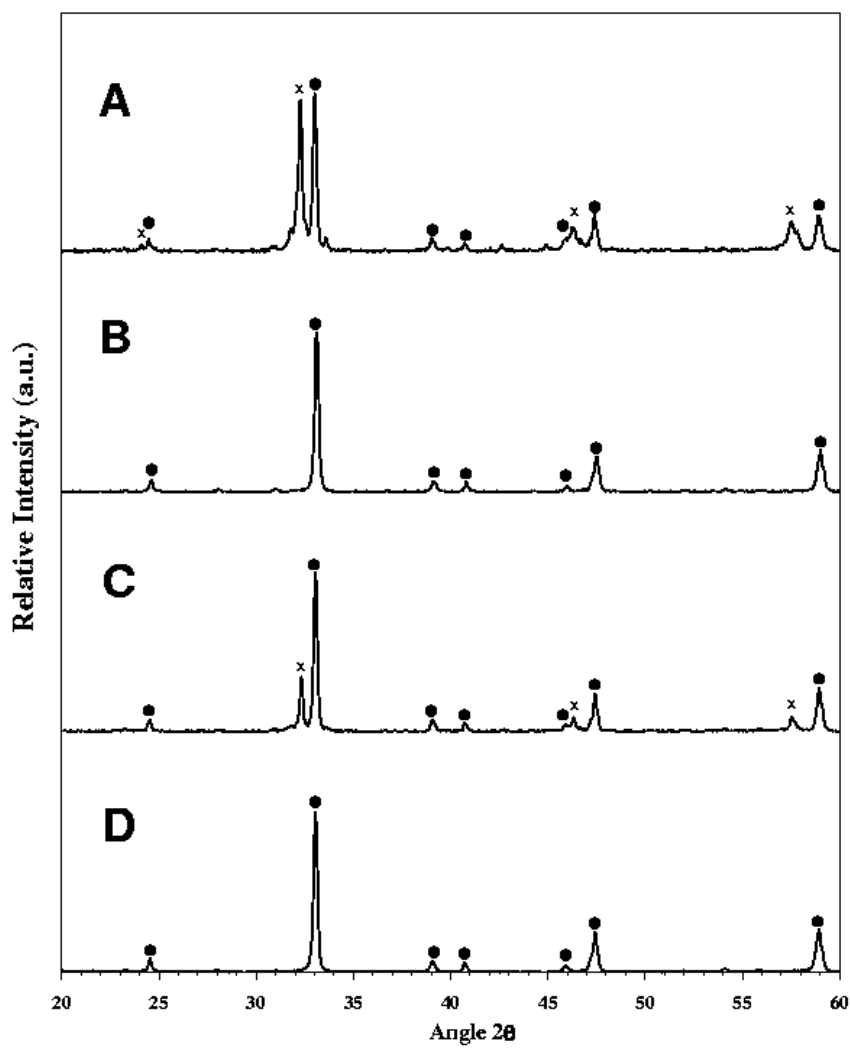

**Fig. 2**: X-ray spectra of two samples of $RuSr_2GdCu_2O_8$ synthesized at 960 °C (A, B) and 1000 °C (C, D). Spectra A and C were taken after the calcination step. B and D show the final spectra after long time sintering at 1065 °C. The reflections marked by a dot indicate the 1212 structure. The crosses show the presence of $SrRuO_3$ after the starting materials were calcined.

The synthesis of other superconducting ferromagnets, e.g. $RuSr_2EuCu_2O_8$ [7] or $RuSr_2(Gd/Ce)_2Cu_2O_{10-\delta}$ [1, 2, 29], is very similar to the procedure described above. Preparing the Ru-1222 structure requires slightly higher temperatures with good results obtained if the final annealing was done at about 1090 °C [29] although the conditions reported from different groups vary slightly.

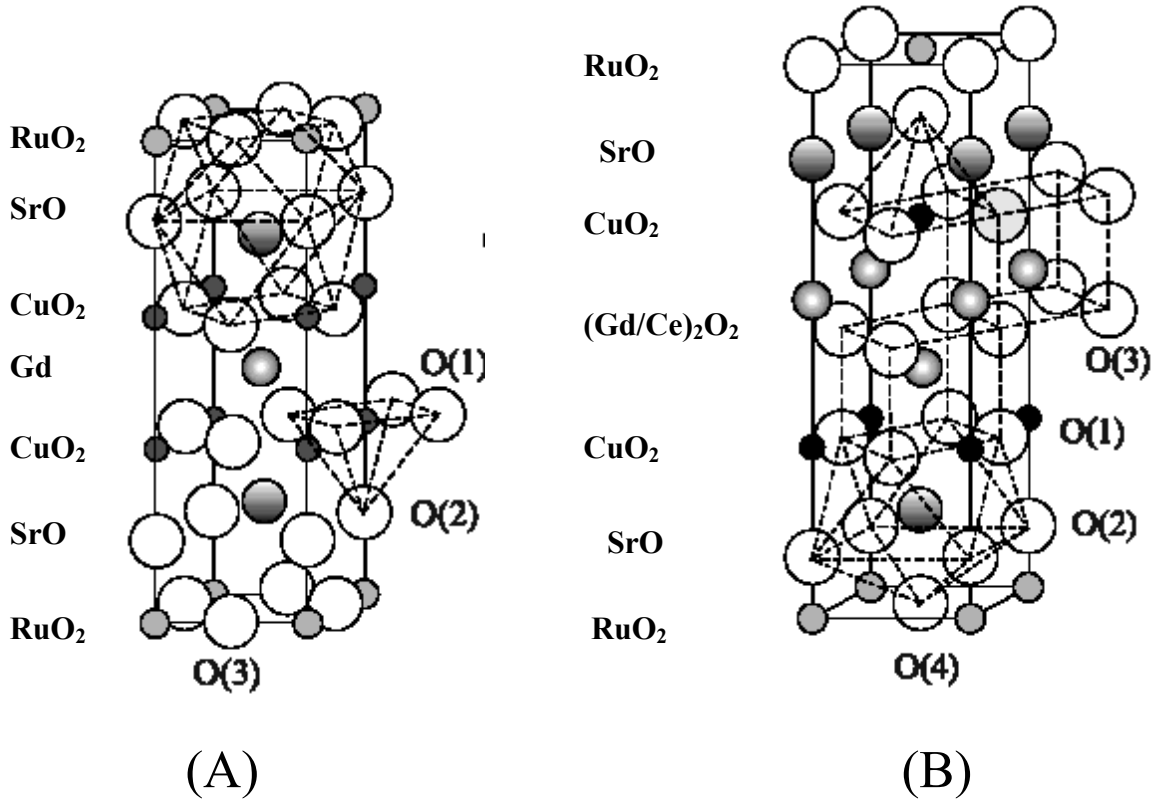

**Fig. 3**: Lattice structure of     **(A)** $RuSr_2GdCu_2O_8$ (Ru-1212) and
                                            **(B)** $RuSr_2(Gd/Ce)_2Cu_2O_{10}$ (Ru-1222)

The lattice structure of both Ru-1212 and Ru-1222 is schematically shown in Fig. 3. Both compounds crystallize in tetragonal symmetry with space groups P4/mmm (Ru-1212) and I4/mmm (Ru-1222), respectively [1, 2, 26, 36, 37]. The structure of Ru-1212 (Fig. 3 A) may be derived from the well-known lattice of $YBa_2Cu_3O_{7-\delta}$ (YBCO) by the following replacement: Y $\Rightarrow$ Gd, BaO layer $\Rightarrow$ SrO layer, and CuO chains $\Rightarrow$ $RuO_2$ planes. In contrast to YBCO the $RuO_2$ layers are fully occupied by oxygen resulting in a change of symmetry from orthorhombic (YBCO) to tetragonal (Ru-1212). The Ru ion is octahedrally coordinated and the $RuO_6$ octahedra deviate from their ideal positions by a small rotation (about 13 °) around the c-axis and a slight tilting that reduces the Cu-O-Ru angle to 173 ° [36]. The c-axis lattice parameter (c=11.5731 Å) is about three times larger than the a-axis parameter (a=3.8348 Å). The almost perfect match of a=c/3 gives rise to the formation of mcirodomains, and the c-axis of adjacent domains may be aligned in different principal directions [26, 36].

The Ru-1222 structure (Fig. 3 B) is derived from the Ru-1212 structure in replacing the Gd ion by a three-layer fluorite-type $(Gd/Ce)_2O_2$-block that causes a shift of the subsequent perovskite block along [110] direction and doubles the unit cell along the c-axis. The tetragonal lattice constants are of the order of a=3.85 Å and c=28.7 Å [1, 2]. The structure and lattice parameters have been confirmed and refined very recently and a tilt and rotation of the $RuO_6$ octahedra similar to that discussed for Ru-1212 was reported [38].

3. MAGNETISM IN Ru-1212 AND Ru-1222

There is still no common agreement on the magnetic structure of these rutheno-cuprates, although almost all possible probes, i.e. neutron powder diffraction (NPD) [16, 26, 39], nuclear magnetic resonance NMR (in particular, zero-field NMR, ZFNMR) [28, 40], macroscopic magnetizations [22, 41, 42], ferromagnetic resonance (FMR) [43], muon-spin rotation (μSR) [22], and Mössbauer spectroscopy (MS) [3], have been used. It is particularly striking that the observations of different probes often were interpreted as mutually exclusive structures when the results of the same technique merge remarkably well. This should be an indication that our interpretation may need to be modified, but most of the experimental data are valid. The issue is particularly important because it may decide how SC coexists with FM, as will be discussed in the next section.

NPD typically is the most powerful tool in determining long-range average spin-structures of homogeneous magnets. The aligned magnetic moments interact with the magnetic dipole of neutrons, which leads to extra magnetic scattering on the top of the usual nuclear scatterings. For example, an AFM spin-order in a particular crystalline orientation may double the unit cell in that direction. Super-lattice like lines, therefore, appear with corresponding Miller indices. An FM ordering, on the other hand, alters the scattering amplitudes of neutrons by adding a magnetic contribution, which can be identified by the change of the line-intensities across $T_m$. The orientation of the ordered moments can be further extracted from intensity ratios of these magnetic lines based on the dipole interactions between neutrons and electrons. The data interpretation, therefore, is relatively straightforward and reliable for the average structure of localized moments. However, it has limitations. In particular, randomly dispersed short-range ordering can be missed due to the line-broadening caused by the uncertainty principle $\Delta\Theta = h/kd$, there $h$, $k$, and $d$ are the Planck constant, the neutron momentum, and the spatial extension of the spin-order, respectively.

In contrast, NMR is a local probe and sensitive to the nearest neighbor interactions, although its interpretation is more model-dependent. NMR essentially measures the *rf* transverse susceptibility of the chosen nucleus and allows an estimation of the hyperfine field there, typically caused by the induced wave-function deformation of the inner-shell electrons. Although the data might appear similar in both AFM and FM spin-orderings, it has been demonstrated that the two can be identified through both the signal-enhancement and the line-shift by the external field [44]. On one hand, the external *rf* field acting on a nucleus will be enhanced by the coherent motion of the FM (or canted AFM) moment [28]. Simultaneously, the signal is suppressed by the exchange field in case of an AFM magnet or by the magnetic anisotropy field in case of a bulk FM magnet [44]. As the result the signal of FM magnets are much stronger. For example, the 375 MHz ZFNMR line of the FM clusters is fivefold stronger than the 290 MHz line of the AFM matrix in $La_{0.35}Ca_{0.65}MnO_3$, although the volume fraction of the FM clusters is only 8%. On the other hand, the NMR line of an AFM magnet

(with or without a small canting angle) will only be broadened without a significant shift by external fields, especially in ceramics with random grain orientations. This is in contrast with FM magnets, where a field-induced line-shift is noticeable with typical hyperfine fields around 10-100 T. It is therefore possible to distinguish between AFM- and FM-dominated spin-orders based on the field effects [44]. In addition, the orientation of the ordered moment can be extracted from the interference between the NMR and the quadrupole-splitting caused by the crystalline electric fields, which can be calculated [28, 40].

It is almost impossible to extract the detailed magnetic structure from macroscopic magnetization measurements alone. However, it should serve as the decisive constraint, which all interpretations need to meet. This, for example, can be well demonstrated in the case of $La_{0.35}Ca_{0.65}MnO_3$, where the magnetization shows two-step like drops around 275 K and 160 K, respectively [45]. The two transitions were initially interpreted as charge-order and AFM transitions, respectively. The rather large low-T magnetization, M/H ≈ 0.01 emu/mole at 100 K, further suggests that either a spin canting or the coexistence of FM species may occur below the AFM transition. A NPD investigation, however, observed no evidences for long-range FM correlations [46]. Detailed ZFNMR data finally revealed that 8% of the compound should be in a FM state [44]. The small size of the FM species was used to justify the difference between the NPD and the ZFNMR data [44].

Similar contrasts are even more drastic in the rutheno-cuprates. Even in the relatively simple case of Ru-1212, where the spin alignment from both NPD and NMR seems to fit the mean-field theory rather well below a well-defined $T_m$ ≈ 135 K, controversies still persist. Although a homogeneous spin-canted model was proposed to accommodate the G-type AFM structure suggested by the (*1/2, 1/2, l/2*) lines of NPD and the significant spontaneous moment (≈ 0.28 $\mu_B$/Ru for Ru-1212-Gd at 5 K) derived from the macroscopic magnetization, both the spin orientation and the magnitude of FM components are controversial. By comparing the (*00l*) line intensities below and above $T_m$, three independent NPD investigations concluded that the possible FM component is < 0.3 and < 0.1 $\mu_B$/Ru for Ru-1212-Gd and 0.34±0.1 $\mu_B$/Ru for Ru-1212-Y at H = 0. Lynn et al. further suggested that the Ru contribution to the FM component at 80 K will be kept below 0.2 $\mu_B$/Ru with H up to 7 T by subtracting the expected paramagnetic Gd contribution. These rather strict limits are even in disagreement with the macroscopic magnetizations. In fact, the spontaneous magnetization, $M_r$, of a ceramic Ru-1212-Gd sample was reported to be 800 emu/mole and corresponded to a FM component of 0.28 $\mu_B$/Ru at 5 K and H = 0 after a geometric correction for the random distribution of the magnetic easy-axis, *i.e.* the orientation of the grains. This is noticeably larger than the upper limit of 0.1 $\mu_B$/Ru at H = 0 by NPD on the sample from the same group. Although the FM-component at higher fields is masked by the paramagnetic contribution of Gd, it is not unreasonable to estimate the value by extrapolating the linear high-field magnetization to H = 0 since the paramagnetic part should be linear and the canted FM-component of Ru may be more-or-less H-independent (as suggested by the NPD data above 1 T). This extrapolated moment of the Ru-1212-Gd sample, ≈ 4000 emu/mole (0.6 $\mu_B$/Ru) at 50 K, however, is again several times higher than the upper limit set by the NPD data. It is interesting to note that the rather similar high-field magnetizations of Ru-1212-Gd and Ru-1212-Eu, *i.e.* ≈ 4000-5000 emu/mole at 50 K and 5 T, suggest that the overall Gd contribution to the aligned moment should be relatively minor up to 5 T. This is again in disagreement with the NPD interpretation reported.

Some other magnetization data even directly challenge the G-type AFM ordering (with or without canting) suggested by NPD. Butera *et al.* summarized: "it seems to be incompatible with magnetization measurements in several aspects: *i)* We have found a *positive* Curie-Weiss constant smaller than the ordering temperature indicating that the predominant interaction among Ru ions is FM. The G-type structure, in fact, requires AF interactions. *ii)* The low-temperature magnetization, which is close to the saturation value, is not consistent with a small canting of the AF alignment. *iii)* EPR results show that there is a net magnetic field of ≈ 600 Oe at the Gd site coming from the ordered array of Ru, whereas a cancellation of the Gd-Ru interaction is predicted in a G-type-ordered AF."

Some of the FM aligned Ru-spins, which should add up to the macroscopic magnetization observed, seem to be missing in the NPD. The ZFNMR data appear to support this speculation. Both a signal-enhancement by a factor of about 100 and a shift to lower frequency with the external field were reported in a Ru-1212-Y sample [28]. Similar enhancements have also been observed in Ru-1212-Gd [28]. The authors, therefore, urged to conduct new detailed NPD experiments [40].

The situation about the orientation of the ordered Ru-spins is similarly confusing. The NPD data suggest that the Ru moments are ordered along the *c* axis based on the intensity ratio between {*1/2, 1/2, 1/2*} and {*1/2, 1/2, 3/2*} lines, *i.e.* with the observed ratio of 2.5(4) compared to a calculated value of 2.2 in this direction. Both ZFNMR experiments [28, 40], however, show that the moments should be aligned within the $RuO_2$ plane. The same conclusion has been reached by the FMR investigation of Ru-1212-Gd. The rather large magnetic anisotropy field of $H_z \approx 11$ T from FMR is in contrast with the much smaller threshold of 0.4 T for the spin-flop observed in NPD [43]. The same question of whether NPD and NMR/magnetizations/FMR probe the same magnetic correlations, therefore, can be raised again.

The magnetic structure of Ru-1222 is even more complicated. There is apparently more than one magnetic transition above $T_c$. Whereas the main low-H transition, defined by the large step-like increase of the field-cooled magnetization ($M_{FC}$) with the sample cooling at H < 0.1 T, typically occurs below 100 K, an AFM-like transition at a much higher temperature of $T_1 \approx 160$-250 K is clearly detected in the high field differential susceptibility (Fig. 4) [42, 47]. The interpretation that the bulk part of Ru-1222 is in an ordered AFM state seems to be supported by the MS data [3]. The MS of trace $^{57}Fe$ in a $RuSr_2Gd_{1.4}Ce_{0.6}CuO_{10}$ sample close to $T_1 \approx 180$ K was interpreted as the result of a hyperfine field of 467 kOe at the Ru site, *i.e.* the Ru moments are ordered up to ≈ $2T_m$. Moreover, two additional minor $M_{FC}$ steps appear in almost all reported magnetization data around 120 and 160 K, respectively [42]. Either several sequential AFM/FM transitions or nanoscale phase-separation should occur. In particular, the wide separation between low-H FM-like transition at $T_m$ and the AFM-like transition at $T_1$ makes a simple canted AFM model less likely [42, 48].

It should be noticed that several data sets have been used to argue that the magnetic order in the rutheno-cuprates is homogeneous on a microscopic scale. One is the widely quoted ZF μSR data of Bernhard et al. [22]. However, the resolution of this experiment is rather limited. The random orientation of the grains (therefore, the random angle θ between the spontaneous magnetization and the muon spin) makes the internal field deducable only as an average. The authors, therefore, estimated that the resolution for the possible magnetization inhomogeneity <ΔB>/B is no better than 20% with B ≈ 700 Oe. The diamagnetic field of a homogeneous FM magnet with a spontaneous magnetization of 800 emu/mole would be well within this range of uncertainty. In addition, the estimated resolution for minor magnetic species is only 20%. The possible phase separation can hardly be excluded by the μSR data, in our opinion. Another suggested evidence for magnetic homogeneity is the temperature dependence of the spin-lattice relaxation rate $1/T_1$ of the $^{101}$Ru ZFNMR line. The relaxation seems to be similar to that of $^{63}$Cu at 10 T, which should be dominated by the formation of the superconducting gap [28]. The authors, therefore, concluded that the aligned Ru-spins and the supercarriers in the $CuO_2$ planes do coexist on a microscopic scale. The coexistence, however, is not necessarily the same as a homogeneous magnetic state. The Josephson-junction like superconductivity observed in these rutheno-cuprates suggests that the FM domains should have a dimension comparable to the proximity length of Cooper pairs if they serve as the tunneling junctions [35]. This geometric proximity may make the $1/T_1$ of $^{63}$Cu and $^{101}$Ru look similar, but still belong to different magnetic species.

To accommodate these apparent discrepancies in the reported data, we investigated the magnetizations of both Ru1212 and Ru1222. Our data suggest that a mesoscopic phase-separation between FM and AFM species occurs in both Ru-1212 and Ru-1222.

### 3.1    Separable and Tunable AFM and FM Transitions

The zero-field-cool magnetization ($M_{ZFC}$) and field-cool one ($M_{FC}$) at 5 Oe of a Ru-1222-Eu sample (A) are shown in Fig 4. The $T_m$ appears around 65 K with a transition width around 20 K. Two additional transitions can be noticed at 120 and 150 K, respectively. To explore the magnetic signature of the AFM transition, the contributions from the Eu and $CuO_2$ are subtracted from the differential susceptibility $\chi$ at 5 T and the *dc* magnetization at 1 T according to the procedure proposed by Butera et al. (Fig. 4) [41]. Both are in agreement above 200 K, and a Currie-Weiss (CW) fit has been made in this temperature range with the deduced Currie temperature and the Ru-moment of 80 K and 2.6 $\mu B$/Ru, respectively. The calculated differential $1/\chi_{Ru}$ follows the extrapolated Currie-Weiss (CW) fit (the solid line) rather well above 150 K. It, however, deviates to the higher value, *i.e.* a smaller magnetization, at lower-T and finally bends around $T_1 \approx 120$ K, a temperature still far above $T_m$ (Fig. 4). However, the *dc* magnetization at 1 T is higher than the CW fit below 200 K. This non-linear H-dependence, in our opinion, is an indication of the existence of nano-domains with FM correlations, a phenomenon widely discussed in CMR compounds [44]. It should be noticed that the suppression of the susceptibility caused by the aligned FM species is relatively small and cannot count for the AFM-like minimum. For example, the *dc* magnetization at 5 T and 100 K is only $\approx 0.3$ $\mu B$/Ru, far smaller than the estimated Ru paramagnetic contribution of 2.6 $\mu B$/Ru. In contrast, the value of $T \cdot \chi$, *i.e.* a parameter proportional to the number of the paramagnetic spins, is only a quarter of that expected from the CW fit at 100 K. AFM spin-spin correlations may be the only reasonable interpretation to count for such large suppression, which is supported by the large hyperfine field observed in MS spectra [3]. The AFM and FM transitions, therefore, seem to occur at very different temperatures in Ru-1222, which challenge the simple canted AFM models proposed.

It is interesting to note that a noticeable extrapolated zero-field magnetization appears even above $T_1$, *e.g.* up to $T_1' = 140$ K in the Ru-1212-Eu sample with $T_1 = 120$ K (Fig 5). This non-zero magnetization and the associated non-linear isothermal M(H), as will be discussed below, can be easily understood as embedded superparamagnetic species. The two additional transitions in the low-field $M_{FC}$ of Fig. 4 may have the similar origin. Their deposition from the otherwise AFM-correlated matrix reminisces the phase-separation widely observed in CMR, *e.g.* in $La_{0.35}Ca_{0.65}MnO_3$ [44].

Felner et al. has previously identified the AFM transition of Ru-1222 with the non-zero extrapolated zero-field magnetization, our differential susceptibility data seem to provide a reasonable explanation, although the two temperatures are slightly different [48].

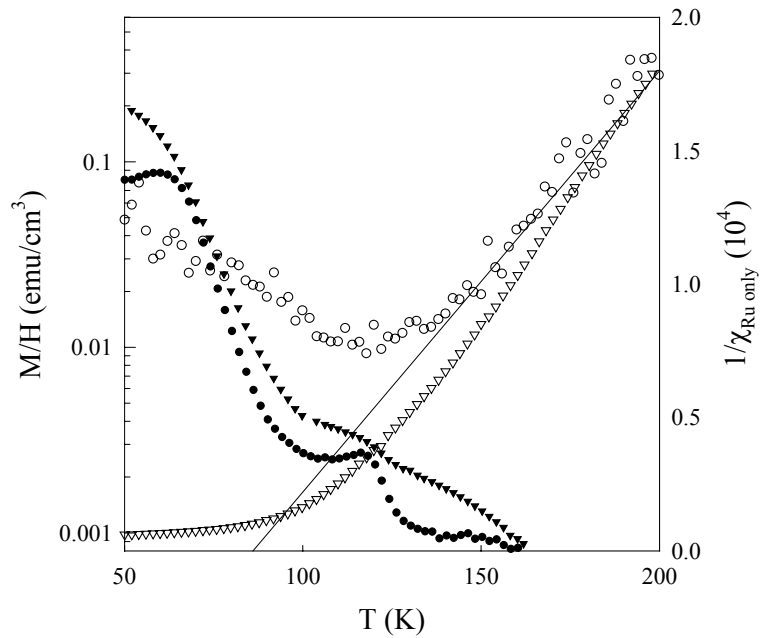

**Fig. 4**: The magnetization of Ru-1222 sample A. ●: ZFC at 5 Oe; ▼: FC at 5 Oe; (the two use the left side scale) ○: the differential susceptibility at 5 T; ∇: dc susceptibility at 1 T (the two use the right side scale).

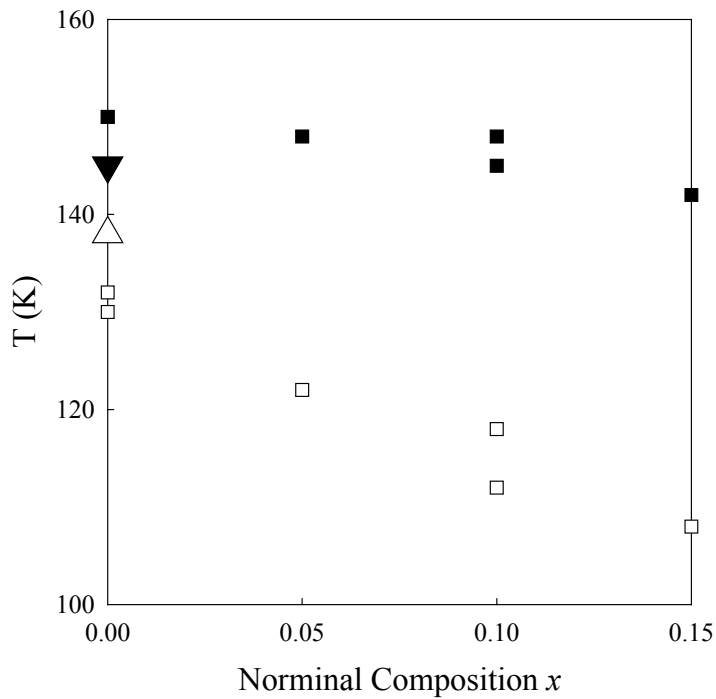

**Fig. 5**: The $T_m$ (□) and $T_1$ (■) of $(Ru_{1-x}Cu_x)Sr_2EuCu_2O_8$ as functions of x. The $T_m$ (△) and $T_1$ (▼) of the Ru1212Eu sample reported in Ref. [41] are also included.

It should be pointed out that a similar situation is observed in Ru-1212, however, the temperature difference between $T_m$ and $T_1$ is far smaller than in Ru-1222. The rough agreement between the $T_m \approx 138$ K and the $T_1 \approx 145$ K reported in a Ru-1212-Eu sample may only be a coincident [41]. Several $(Ru_{1-x}Cu_x)Sr_2Eu_{1.4}Cu_2O_8$ samples were measured by us to explore the issue. The $T_m$ decreases with the x rather quickly (about 20 K with 0.1 Ru-replacement) (Fig. 5). The $T_1$, however, is insensitive to this doping with no more than 5 K decrease up to x = 0.1. The wide separation between the two transitions, therefore, exists even in Ru-1212 if the ferromagnetic interaction has been slightly suppressed (as evidenced by the lower $T_m$). It is also interesting to note that this correlation between $T_m$ and $T_1$ under the Cu-Ru replacement of Ru-1212 is rather similar to that in Ru-1222-Eu with a Ce-Eu replacement, specially considering that $T_1'$ and $T_1$ are used for Ru-1222 and Ru-1212, respectively (Fig. 6) [48]. Ru-1212 and Ru-1222 seem to share the same basic magnetic structure, although the competing magnetic interactions seem to tend more to FM in Ru-1212.

Only with such multistage transitions and possible phase-separation, in our opinion, the extremely broad $C_p$ anomaly on the higher-T side in Ru-1212-Gd can be naturally accommodated with the narrow mean-field-like transition below $T_m$ observed in NPD and NMR.

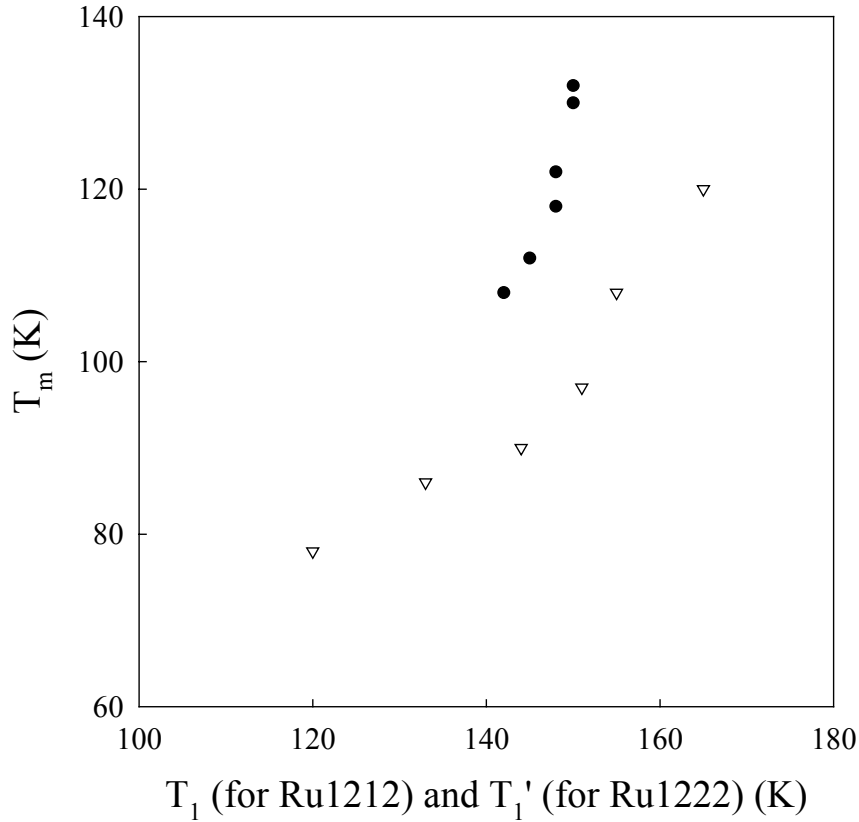

**Fig. 6**: $T_m$ vs. $T_1$ of: the Cu-doped Ru-1212-Eu (●, using the differential susceptibility to determine $T_1$) and Ce-doped Ru-1222 of Ref. [48] (▽, using the non-zero extrapolated zero-field-magnetization to determine $T_1$)

## 3.2   Superparamagnetism in Ru-1212/Ru-1222 far Above $T_m$

To further explore the magnetic states of these rutheno-cuprates, the isothermal M(H) was measured for Ru-1222-Eu (Sample A, Fig. 7) and a Ru-1212-Eu sample (Sample B, Fig. 8a, 8b). The reversible non-linear M(H) is evident up to $T_1$ or higher. Although H-dependent M(H)'s are a common phenomenon in magnets near $T_m$ due to spin fluctuation, its appearance up to temperatures as high as $2*T_m$ is highly unusual. Spin fluctuations, H-induced rotation of the canting angle, and phase separations have been invoked to interpret the data [41, 42, 48, 49]. Regardless of the microscopic mechanism (except for rotation of the canting angle of an AFM magnet), however, the spin-alignment under fields can always be treated as a competition between the thermal energy and magnetic energy.

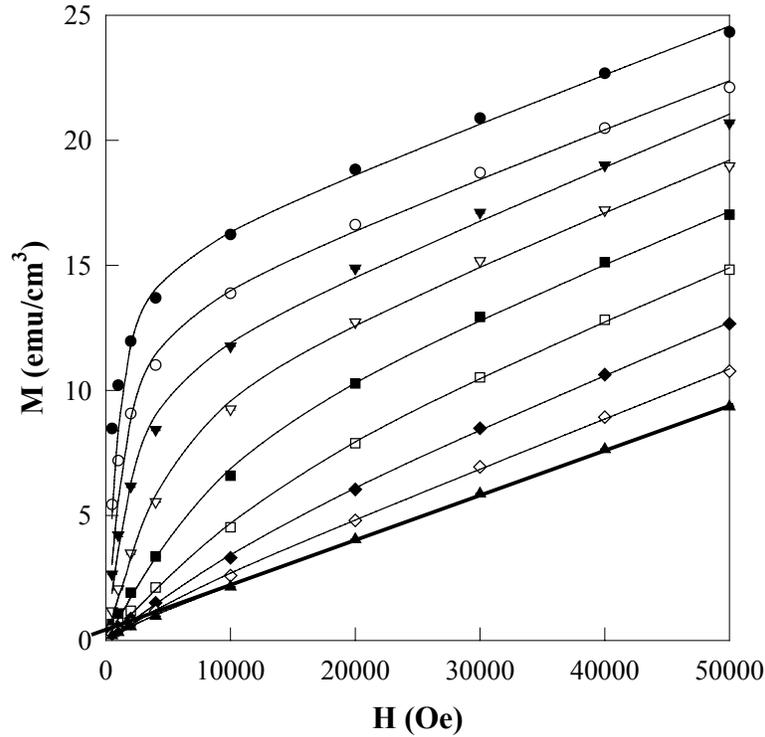

**Fig 7**:   The isothermal M(H) of the Ru-1222-Eu sample A. From top to bottom: 60, 70, 80, 90, 100, 110, 120 130 and 140 K. The bottom thick line shows the non-linearity existing even at 140 K.

This competition can be approximately described by the Langevin function,

$$M = m_0 \left[ \mathrm{ctnh}(\mu H/k_B T) - k_B T/\mu H \right]$$

if the magnetic correlations can be truncated at some distance *l* and the FM spins are treated as isolated clusters (*i.e.* the interaction is much stronger than $k_B T$ within *l* but much weaker at longer distance), $m_0$, $\mu$ and $k_B$ are the saturation moment, the average moment of individual clusters and the Boltzmann constant, respectively. It has been demonstrated that the deviations caused by the truncation, *e.g.* the interactions smoothly scaled with *l* in a second order transition, only underestimate the $\mu$ [50]. Typical spin-fluctuations and phase-separation, therefore, can be identified by the value of μ, which should involve no more than the nearest neighbors in the case of fluctuations far above $T_m$.

In the case of Ru-1212/Ru-1222, an additional linear term should be added to account for the paramagnetic contribution of Gd. It is interesting that the fits (solid lines) are so good for both Ru-1212 and Ru-1222 over a broad temperature range (Figs. 8a, 8b). It is more remarkable that the estimated size of the correlated moments is about 300-2000 $\mu_B$ far above $T_m$, corresponding to magnetically correlated regions of 20-50 Å size assuming a moment of 1 $\mu_B$/Ru. This is decisively different from the case of Ni, a classical sample of critical fluctuation. The cluster size is comparable to the lattice parameters above 1.1 $T_m$ there (Fig. 9).

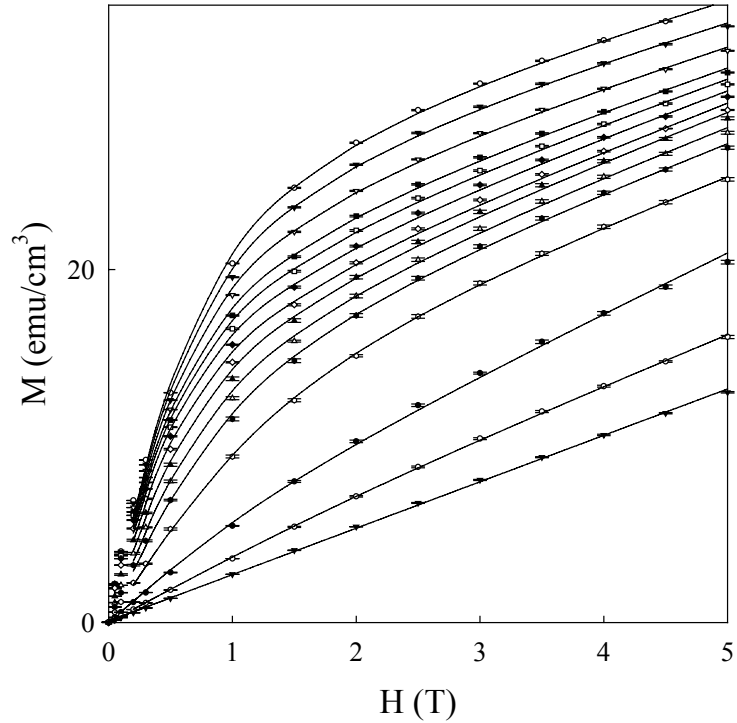

**Fig. 8a**: The isothermal M(H) of Ru-1212-Eu sample B with $T_m$ = 130 K and $T_1$ =160 K. Symbols: data, from the top to bottom: 110, 115, 120, 124, 126, 128, 130, 132, 134, 136, 140, 150, 160 and 170 K. The lines are the fits of Langevin functions.

The layered structure and the presumed 2D FM in Ru-1212/Ru-1222 have been often used to argue in favor of the critical fluctuation model. We disagree. Spin-fluctuations are essentially a competition between $k_BT$ and the magnetic energy, which is proportional to the spins coherently bound. The number of the bound spins, however, will be $(\xi/a)^3$ in 3D but a much smaller value of $(\xi/a)^2$ in 2D, there $a$ and $\xi$ are the lattice parameter and the correlation length, respectively. This is the root for the stronger fluctuation in 2D below $T_m$. Far above $T_m$, however, the smaller coherent volume in a 2D magnet can only make the M(H) more linear. In a second-order transition, in fact, it is widely accepted that the scaling equation can be written in a dimensionless form,

$$(HM_0/MH_0)^{1/\gamma} = (t-1)+(M/M_0)^{1/\beta}.$$

$t$, $\gamma$, $\beta$, $M_0$ and $H_0$ are the reduced temperature $T/T_m$, two critical exponents and two critical amplitudes, respectively. The scaled susceptibility ($MH_0/HM_0$), therefore, is

$$(MH_0/HM_0) = (t-1)^{-\gamma}[1+(H/H_0)^{1/\beta}/(t-1)^{1+\gamma/\beta}]^{-\gamma} \approx (t-1)^{-\gamma}\{1-\gamma[(H/H_0)/(t-1)^{\beta+\gamma}]^{1/\beta}\}$$

in lowest order of $H/H_0$. The H-dependence, hence, can only be observed at an H comparable to $H_0\gamma^{-\beta}(t-1)^{\beta+\gamma}$, and should be suppressed with temperature as $1/(t-1)^{1+\gamma/\beta}$, *i.e.* determined by the critical exponents. The theoretical critical exponents depend on the dimensionality and the interaction range, they are $\beta = 0.5$ and $\gamma = 1.0$ in the mean field theory, $\beta = 0.3$ and $\gamma = 1.3$ in the 3D Heisenberg model, but $\beta = 0.125$ and $\gamma = 1.75$ in the 2D Ising model.

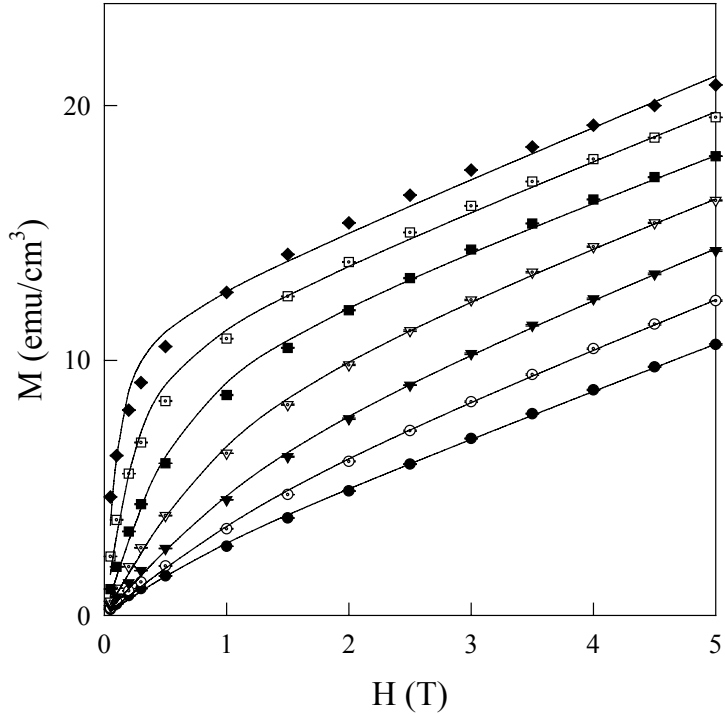

**Fig. 8b**: The isothermal M(H) of Ru-1222-Eu sample C with $T_m = 65$ K and $T_1 = 120$ K. Symbols: data, from the top to bottom: 70, 80, 90, 100, 110, 120 and 130 K. The lines are the fits of Langevin functions.

Experimentally, the values of $1+\gamma/\beta$ range from 3 (some 3D magnets) to > 10 (quasi-2D magnets), *i.e.* $1+\gamma/\beta$ is even larger in 2D magnets. Non-linear M(H) may not be expected as a sole effect of lower dimensionality. An example is the quasi-2D ferromagnet $La_{1.2}Sr_{1.8}Mn_2O_7$. It has been demonstrated that the non-linear M(H) far above $T_m$ in this compound is a result of mesoscopic phase-separations instead of dimensionality [51].

To verify the existence of superparamagnetic clusters, the magnetic relaxations were explored far above $T_m$. The field was increased to 5 Oe after the sample being zero-field-cooled to the given temperature. A logarithmic increase of the magnetization with time was observed in both Ru-1212 and Ru-1222 above $T_m$. The time-dependent magnetization of a Ru-1222 sample, Sample C, with $T_m \approx 65$ K, for example, is shown in the inset of Fig 10, and the decay rate in the main figure.

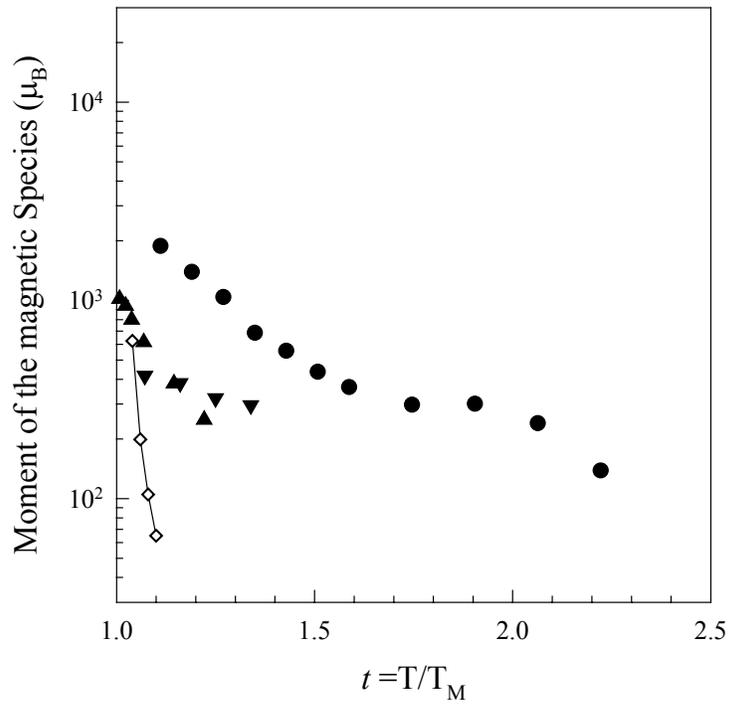

**Fig. 9**: The deduced cluster size of: •: Ru-1222-Eu sample A; ▲ and ▼: Ru-1212 samples B and B'; ◊: Ni.

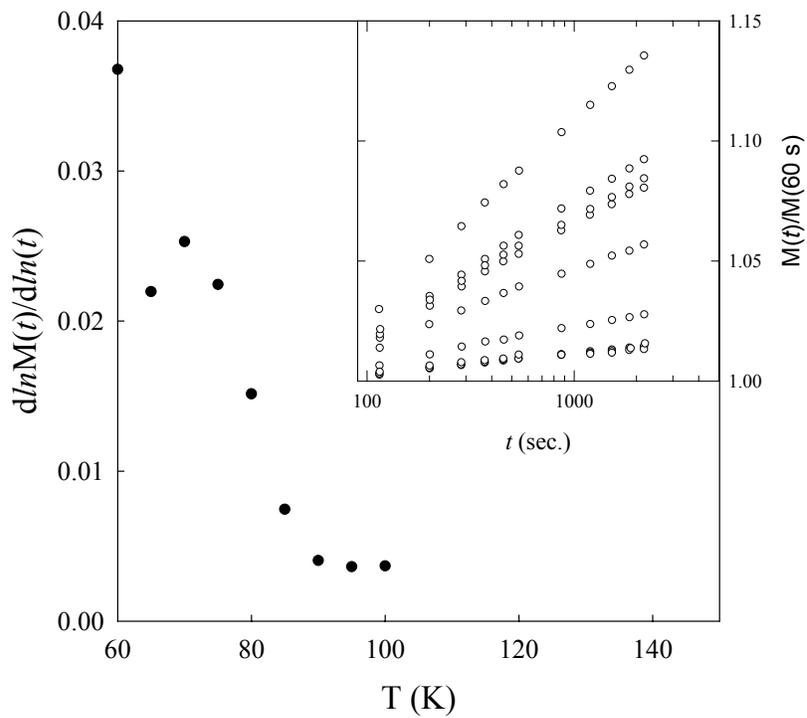

**Fig. 10**: The relaxation rate of $M_{FC}$ at 5 Oe of Ru-1222-Eu sample C. Inset: the raw data.

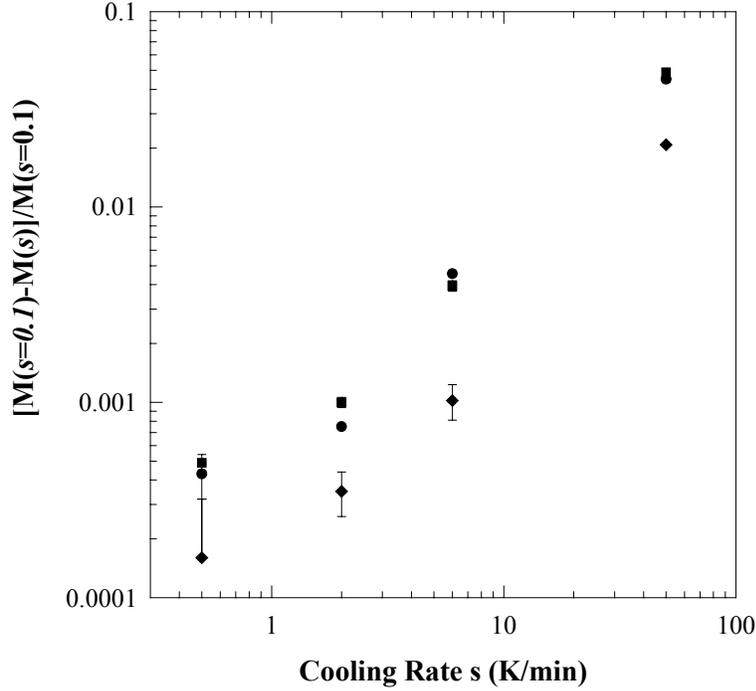

**Fig. 11**: The change of $M_{FC}(60\ K)$ of Ru-1212-Gd sample D at various fields. ■: 0.53 Oe; ●: 0.3 Oe; ◆: 3 Oe.

The slow dynamics also appears as a function of the cooling rate of the samples. A Ru-1212-Gd sample (Sample D), for example, was cooled at given fields from room temperature to 60 K with a rate ranging from 0.5 to ≈ 50 K/min (by dropping the sample directly into the pre-cooled chamber). The sample magnetizations were then measured 30 minutes after the temperature being stabilized at 60 K. A decrease of the $M_{FC}$ with the increase of the cooling rate was observed (Fig. 11). The decrease can be up to 6 %, *i.e.* a significant amount of the spins can be aligned only after 10-100 sec close to $T_m$. The smaller dependence at higher fields again suggests that the cluster size is the origin of the slow dynamics.

Combining the slow dynamics with the non-linear isothermal M(H), we estimated that 1-10 % of the sample is in the form of superparamagntic clusters above 70 K. This is a strong support for the phase-separation models.

### 3.2   Evidences for a Spatial Correlation Between FM and AFM species

To explore the spatial correlation between these different magnetic species, we measured the hysteresis caused by the demagnetizating fields of the 120 K/150 K FM species in Ru-1222 [42]. It was observed that two additional step-like jumps and a main one appear in the $M_{FC}$ at 150 K, 120 K, and at $T_m = 85$ K, respectively, in a Ru-1222-Gd sample, Sample E. This is rather similar to that of Fig. 4. The 85 K FM species should nucleate, therefore, under the combination of the external field H and the demagnetizing fields of the 120 K/150 K species. The combined field, in turn, can be estimated by the difference, $\Delta M_{FC}$, between the

$M_{FC}$'s at 95 K and 60 K, since our early data demonstrate that $M_{FC}/H$ between 40 K and $T_m$ is H-independent below 5 Oe. To enhance the effect of the demagnetizing fields the following experimental procedure was used:

*A* (field-cooling under a fixed field of 10 Oe from 200 K to 105 K) - *B* (switching field from 10 Oe to a lower field of $-1$ Oe $< H_S < 1$ Oe) - *C* (field-cooling under $H_S$ from 105 K to 60 K) - *D* (raising the temperature under the same $H_S$ to 200 K) - *E* (field-cooling under $H_S$ from 200 K to 60 K).

In the sequence of cooling and heating steps described above a significant hysteresis was observed (Fig. 12). In fact, the $\Delta M_{FC}$ can even be negative under a positive $H_S$.

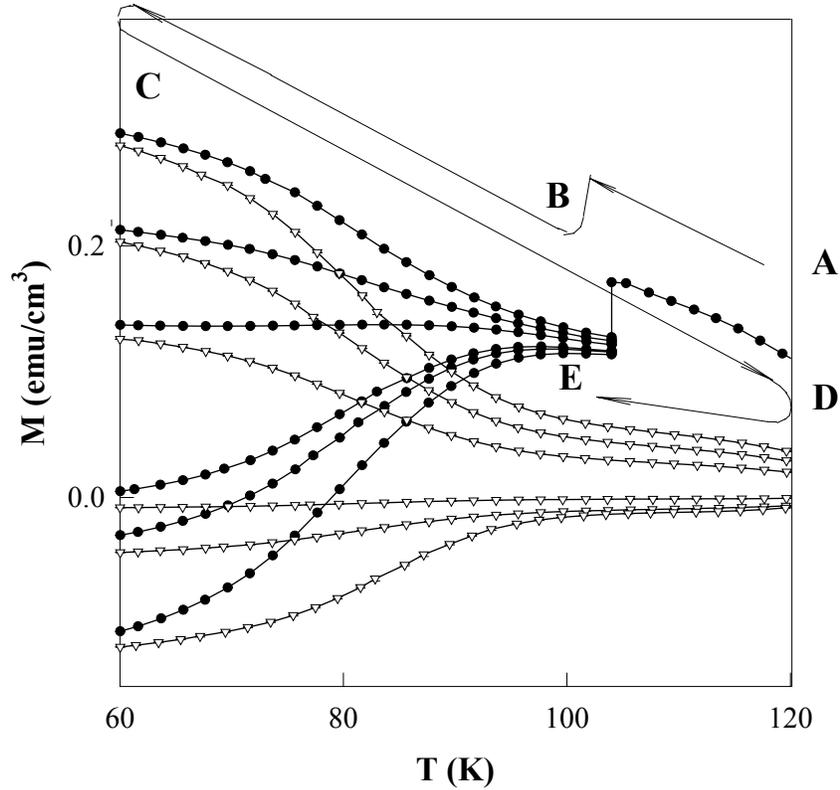

**Fig. 12**: The magnetizations of a Ru-1222-Gd sample E with $T_m = 85$ K in procedure ***A-B-C-D-E*** at various $H_S$ (see text).

The $\Delta M_{FC}$ in step *E*, *i.e.* under a simple field cooling, is the linear function of $H_S$ through the origin, as expected (Inset, Fig. 13). The $\Delta M_{FC}$ in the procedure *C*, however, intercepts with the H-axis at finite value ($H_d \approx 0.6$ Oe in this case) although still being a linear function of $H_S$ (Inset, Fig. 13). It is only natural, therefore, to interpret the $H_d$ as the demagnetizing field of the 120 K/150 K species [with their magnetizations taken as $M_{FC}(95\ K)$]. It is remarkable that all of the measured $\Delta M_{FC}$, *i.e.* in both procedures *C* and *E*, fit onto the same straight line of $\Delta M_{FC} \propto H\text{-}f\cdot 4\pi M_{FC}(95\ K)$ with *f* being an effective demagnetizating factor.

The demagnetization field will be independent of the Ru-1222 particle size of diluted powders when the particle size is larger than the average distance between the adjacent FM species. It, however, should disappear when the particle size is smaller. Several powders with particle size down to 0.8 μm were made from Sample E, and the deduced demagnetizing fields of the 120 K/150 K species field-cooled under 10 Oe field are found to be size-independent. This demonstrates that all these FM species are distributed uniformly at least down to the length scale of 0.8 μm, although the crystalline grain size observed under SEM is much larger (2-20 μm for the sample). We, therefore, attribute the FM species to mesoscopic phase-separations.

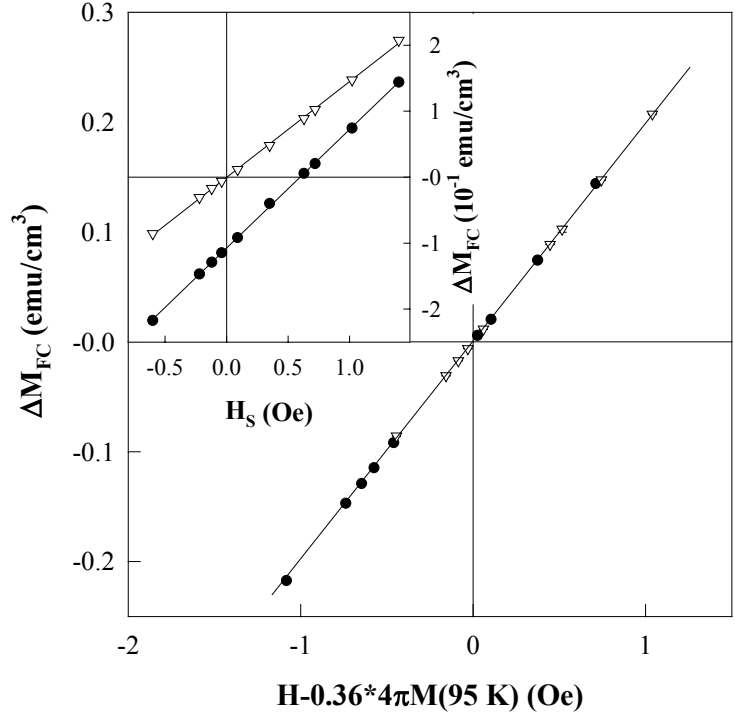

**Fig. 13**: The $\Delta M_{FC}$ vs. the combination field H-$f \cdot 4\pi M$(95 K) for sample E. Inset: $\Delta M_{FC}$ vs. H under procedures $E$ (∇) and $C$ (•).

Another important result from the measurement is the universal value of $f \approx 0.36 \pm 0.1$ for all Ru-1222 samples we measured. After a geometric correction for the random orientation of the easy axis of the 120 K/150 K species, the microscopic demagnetizing factor $f_o$, i.e. the ratio between the demagnetizing field "felt" by a 85 K species and the moment of the adjacent 120 K/150 K species, should be $0.12 \pm 0.03$, as supported by our partially aligned powder sample. The small demagnetizing factor, which should be determined by the spatial correlation between the adjacent 120 K/150 K species and the 85 K species, suggests that the three types of species prefer to be aligned along their easy axis. This can occur only if the FM species occupy only a small part of the crystalline grains, a scenario of phase separation again.

Based on the magnetization measurements, therefore, we conclude that the magnetic structure of these rutheno-cuprates has to be inhomogeneous on a microscopic scale.

## 4. SUPERCONDUCTIVITY IN Ru-1212 AND Ru-1222

### 4.1 The Underdoped Nature of the Superconducting State

The existence of superconductivity was first documented by Bauernfeind et al. [1] in the Ru-1222 system based on resistivity and ac susceptibility measurements. The bulk superconductivity in Ru-1212, however, was questioned because of the lack of a clear Meissner signal in field-cooled magnetization experiments in the majority of 1212 compounds investigated by different groups (c.f. the extensive discussion in Section 1). The nature of Cooper pair formation and superconductivity is most likely to be similar to that widely discussed for many high-temperature superconducting cuprate materials, e.g. $YBa_2Cu_3O_{7-\delta}$. In particular, the structural similarity of YBCO and Ru-1212, as discussed in Section 2, is indicative for a similar mechanism of superconductivity in both compounds. More than a decade of research on high-$T_c$ perovskites revealed a characteristic phase diagram and universal relations between normal-state transport properties, the onset of superconductivity ($T_c$), and the doping state (the average hole density $p$ in the $CuO_2$-planes) that are common to all high-$T_c$ systems. Comparing all these characteristic quantities we will see that the Ru-1212 and Ru-1222 cuprates are typical underdoped high-$T_c$ systems.

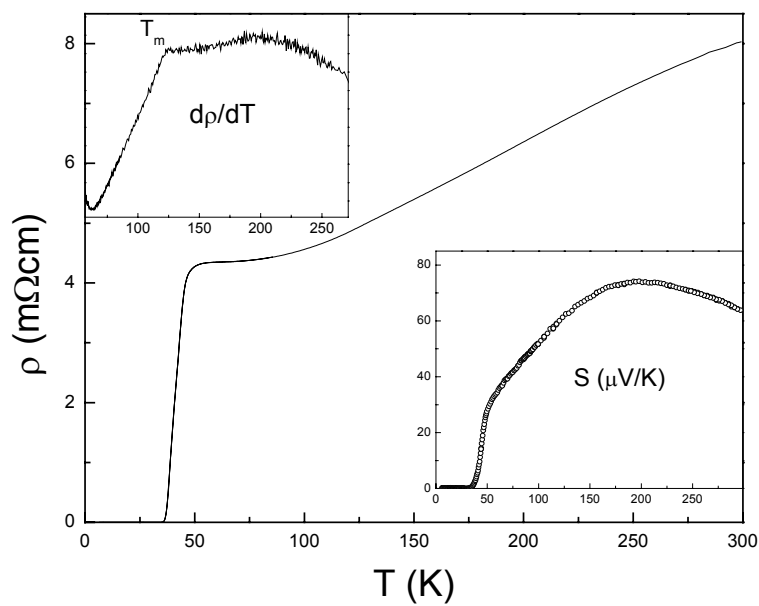

**Fig. 14**: Resistivity and Seebeck coefficient (lower inset) of $RuSr_2GdCu_2O_8$. The upper left inset shows the derivative, $d\rho/dT$, with a clear slope change at the magnetic transition temperature, $T_m$.

The first indication of the underdoped nature of the superconducting state comes from transport measurements in the normal state. The resistivity, $\rho$, of a $RuSr_2GdCu_2O_8$ sample is shown in Fig. 14. The magnetic transition at about 130 K is reflected in a sharp change of slope of the derivative, $d\rho/dT$ (upper left inset). Moreover, the resistivity below room temperature also shows the typical curvature that was observed in other underdoped high-$T_c$ materials and was interpreted as the opening of the pseudo-gap. The strongest evidence for the low carrier density is the high value of the room-temperature Seebeck coefficient (lower inset in Fig. 14). For the RuSr$_2$GdCu$_2$O$_8$ system the value of S(290 K) varies between 70 and 80 µV/K [8, 24] and it was shown that it is almost independent of the conditions of synthesis, annealing etc. [8]. Adapting the empirical relation between S(290 K) and the hole density $p$ that was shown to be valid for most high-T$_c$ cuprates [52],

$$S(290\ K) = 992\ \exp\{-38.1\ p\}\ ,$$

the hole density in the CuO$_2$ plane can be estimated as $p \approx 0.07$. This value of the carrier density for the stoichiometric Ru-1212 appeared to be very rigid and could not be increased by optimizing the conditions of synthesis or by annealing in (high pressure) oxygen [8]. This indicates that in Ru-1212 the oxygen content is very close to 8 and cannot be varied at will to change the hole density (as e.g. in YBCO). This conclusion is also in agreement with the results of a thermogravimetric analysis [5]. The systematic change of carrier density in Ru-1212 may be achieved by cation doping, e.g. by replacing Ru with Cu [53].

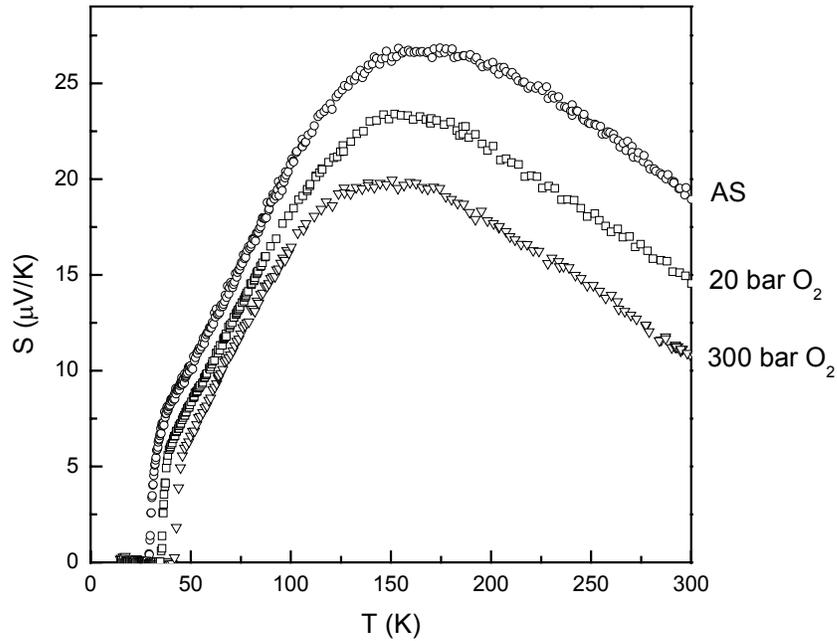

**Fig. 15**: Seebeck coefficient of RuSr$_2$(Gd$_{0.7}$Ce$_{0.3}$)$_2$Cu$_2$O$_{10+\delta}$. The curves show data for the as synthesized as well as for high-pressure oxygen annealed (20 and 300 bar) samples.

Thermoelectric power measurements on RuSr$_2$(Gd,Ce)$_2$Cu$_2$O$_{10+\delta}$ yield typical values at 290 K of about 20 µV/K [29] corresponding to a hole density of $p = 0.1$. This value is distinctly larger than the carrier density in Ru-1212 but is still well below the optimal doping value of 0.16. Unlike the Ru-1212 system, the oxygen content ($\delta$) in Ru-1222 can be modified by high pressure oxygen annealing resulting in an increase of the carrier density. This was clearly demonstrated by measuring the Seebeck coefficient and by measuring the oxygen content directly in gas effusion experiments [29]. As shown in Fig. 15, S(290 K) decreases systematically with annealing of the as synthesized sample in oxygen pressures of 20 bar and 300 bar. However, the change of S(290 K) from 20 µV/K (as) to 11.5 µV/K (300 bar $O_2$) is small corresponding to an increase of the hole density, *p*, by not more than 0.015.

### 4.2 Inter- and Intra-Grain Superconductivity in Superconducting Ferromagnets

The synthesis process described in Section 2 leads to the formation of powder or ceramic pellets of the rutheno-cuprates. Superconductive single crystals of Ru-1212 have been reported but the high density of defects/weaklinks suggested by the small ZFC diamagnetic signal severely limit their advantage [54]. Most granular high-temperature superconductors, however, exhibit a weak inter-grain coupling in the superconducting state resulting in a lower zero resistance temperature and large differences in the FC and ZFC diamagnetic signals. In particular, two well resolved steps in passing through the transition are frequently observed in susceptibility as well as the resistivity data. The onset of superconductivity is usually observed at $T_c$ where the grains become superconducting. Different grains are coupled via the Josephson effect resulting in a phase coherence of the superconducting order parameter across the grain boundaries at a lower temperature $T_p$, the phase-lock temperature.

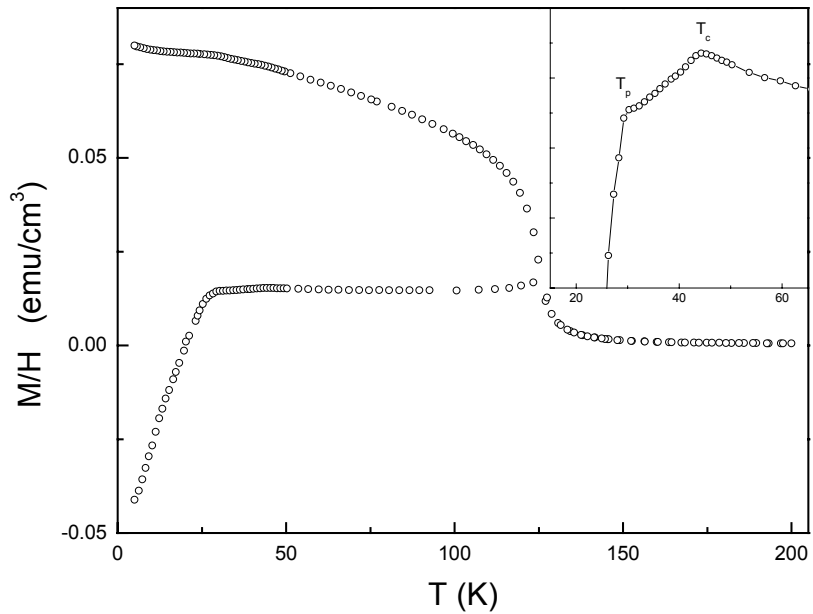

**Fig. 16**: Magnetization data of $RuSr_2GdCu_2O_8$ at an applied field of 7 Oe. FC and ZFC data are shown in the upper and lower branch, respectively. The inset is an enlargement of the ZFC data close to the superconducting transition showing the intra-grain critical temperature ($T_c$) and the inter-grain phase-lock temperature ($T_p$).

These features of a granular superconductor are clearly seen in the Ru-1212/Ru-1222 compounds. Fig. 16 displays a typical example for $RuSr_2GdCu_2O_8$. The ZFC dc susceptibility exhibits a strong diamagnetic drop at about $T_p$=30 K. The inset of Fig. 16 enlarging the data in the vicinity of the transition shows, however, that the diamagnetic decrease of $\chi_{dc}$ actually sets in at a higher $T_c$=44 K. The same two-step like drop is also detected in the real part of the ac-susceptibility. $\chi'_{ac}$ measures the shielding signal of a bulk superconducting sample and should be compared to the ZFC $\chi_{dc}$. Fig. 17 shows the ac susceptibility of another $RuSr_2GdCu_2O_8$ sample with the magnified transition region in the inset. Although the characteristic temperatures are slightly different for this sample ($T_p \approx 36$ K, $T_c \approx 43$ K), the similarity to the ZFC $\chi_{dc}$ data of Fig. 14 is obvious. It should be noted that the intra-grain diamagnetic signal is extremely small for reasons that will be discussed in the next paragraphs. This typical behavior was observed for many different Ru-1212 as well as Ru-1222 samples [7, 8, 29, 55, 56].

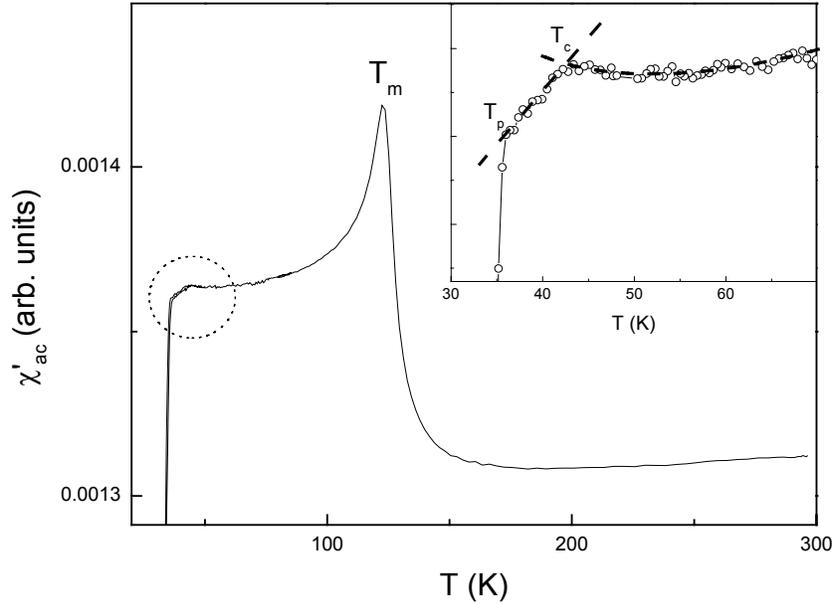

**Fig. 17**: AC susceptibility of a ceramic $RuSr_2GdCu_2O_8$ sample. The peak marks the magnetic transition temperature. The inset shows the details of the inter-grain ($T_p$) and intra-grain ($T_c$) superconducting transitions.

The inter- and intra-grain superconducting transitions are expected to affect the resistivity at the transition since the electrical transport is sensitive to weak links across the grain boundaries. In fact, a broad resistive superconducting transition, sometimes well divided into two steps, has been observed in most transport investigations [7, 8, 34, 53, 55, 57, 58]. One example is shown in Fig. 18. The derivative, $d\rho/dT$, clearly reveals the superposition of two peaks corresponding to the two resistive transitions observed at $T_c$ (intra-grain) and $T_p$ (inter-grain). The magnetic field dependence of the resistivity is used to further verify the interpretation because the weak inter-grain links are supposed to be quickly suppressed by small fields. The resistivity data of Fig. 18 for low magnetic fields of 200 and 500 Oe show the expected dramatic decrease of the inter-grain $T_p$. The zero resistance state is strongly suppressed by the external field. All these observations can well be understood within the simple picture of intra- and inter-grain superconducting transitions in ceramic superconductors.

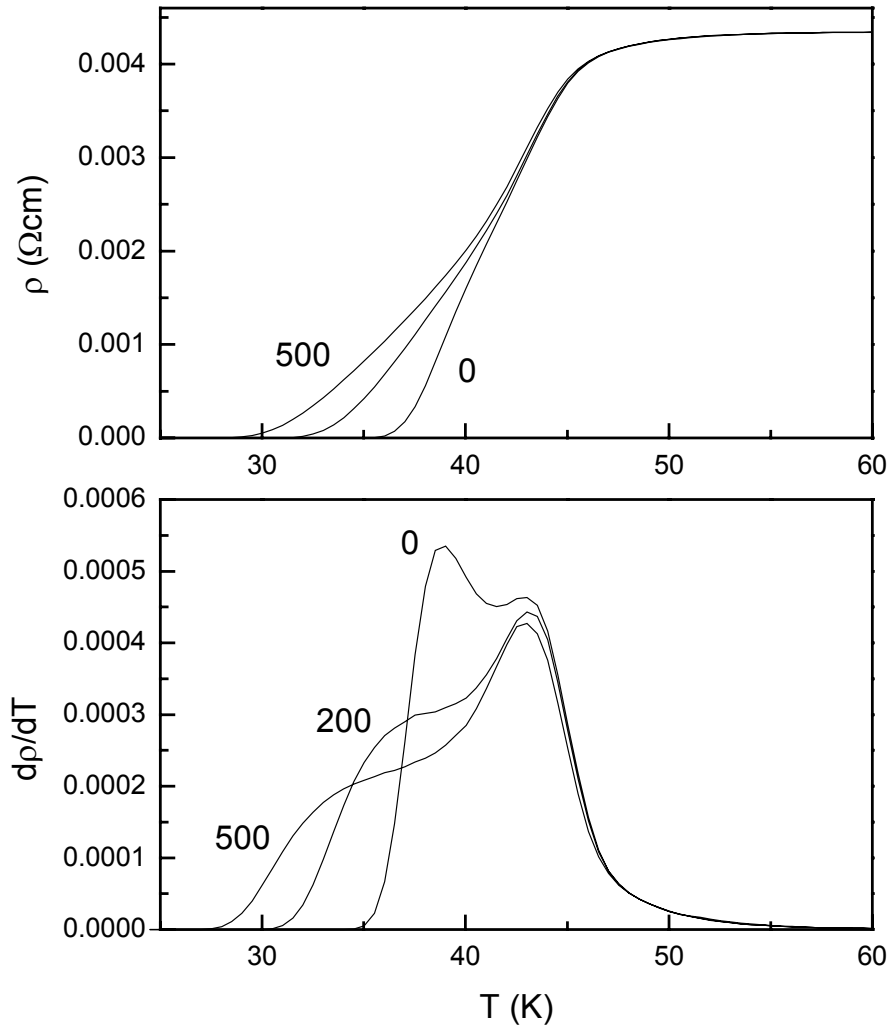

**Fig. 18**: Resistivity of $RuSr_2GdCu_2O_8$ at the superconducting transition. The derivative, $d\rho/dT$, clearly shows the superposition of two peaks corresponding to intra- and inter-grain transitions. The numbers indicate the magnitude of an external magnetic field in Oe applied during the measurement.

To collect additional evidence for our interpretation of $T_p$ and $T_c$ we have conducted measurements of the ac susceptibility of sorted powders of Ru-1212/Ru-1222 samples with well-defined particle size [29, 58]. As discussed above, $\chi_{ac}$ for a bulk granular sample shows two distinct steps (Fig. 17). For powders with particle size well above the grain size the two steps should still exist, but the inter-grain shielding signal will decrease with the particle size. If the particle size of the powder is comparable to or smaller than the grain size of the starting material the inter-grain diamagnetic signal will be completely suppressed but the intra-grain signal is expected to survive. Fig. 19 shows an example of the ac susceptibility for sorted powders of $RuSr_2GdCu_2O_8$. The bulk sample, also shown for comparison, reveals the strongest inter-grain diamagnetic shielding signal. The grain size of this sample was estimated from scanning electron microscopy between 2 and 5 μm. The inter-grain shielding signal is largely reduced for the 40 μm powder and completely suppressed for powders with average particle size of 2.3 and 0.8 μm, i.e. smaller than the original grain size. The intra-grain transition, however, is detected also for the powdered samples with an unaltered $T_c$.

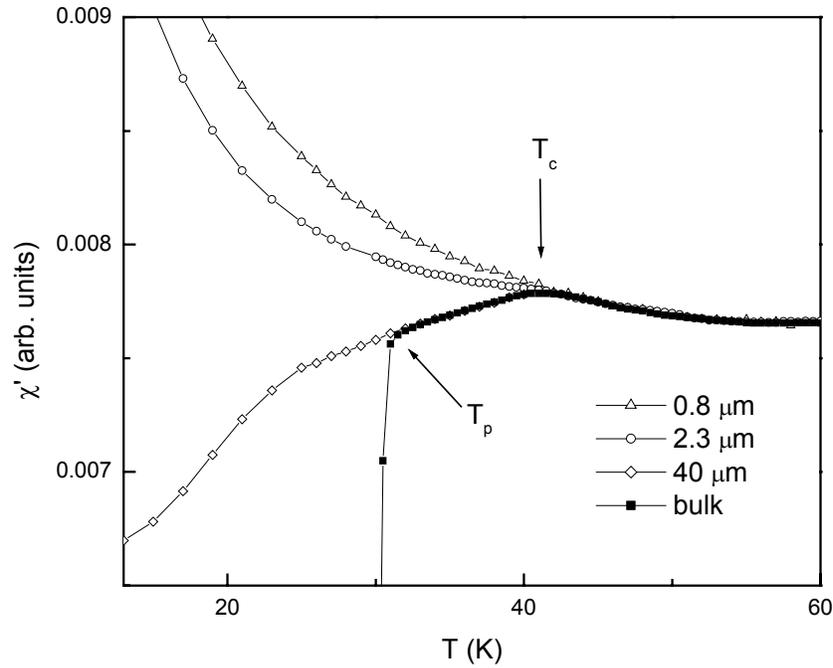

Fig. 19: AC susceptibility of sorted powders of $RuSr_2GdCu_2O_8$. $T_c$ and $T_p$ are easily resolved. The inter-grain transition is completely suppressed for powders with an average particle size (2.3 and 0.8 μm) smaller than the original grain size (2 to 5 μm).

It should be noted that the diamagnetic signals in $\chi_{ac}$ (Fig. 19) appear on top of a paramagnetic background. This paramagnetic signal can be attributed to the magnetic moments of the Gd/Ru ions. It was shown that the Gd moments order antiferromagnetically at a far lower temperature of about 2.8 K [4].

A similar analysis of the diamagnetic shielding signal of bulk and powdered samples on $RuSr_2(Gd_{0.7}Ce_{0.3})_2Cu_2O_{10+\delta}$ led to analogous conclusions for the Ru-1222 system [29]. Apparently, the properties of the superconducting ferromagnets are affected to a large extent by the weakly coupled grains. This explains why in transport measurements the zero resistance temperatures reported by many groups scatter over a broad temperature range. A more detailed analysis how various conditions of sample synthesis and annealing procedures affect the grain-grain coupling as well as the intra-grain superconductivity was given in Ref. [8]. However, the foregoing discussion proofs that inter- and intra-grain transitions can well be separated in both, magnetic and transport measurements. This is a basic preposition to investigate the "intrinsic" properties of the superconducting state. In the following section, therefore, we focus our attention on the intra-grain superconducting properties of the ruthenocuprates.

### 4.4 Effects of Magnetic Phase Separation on the Intra-Grain Superconducting Properties

With the discussion of the magnetic structure and the conclusion of the appearance of phase separation into FM and AFM domains on a sub-grain nanoscopic scale (Section 3) it is expected that the intra-grain superconductivity exhibits exotic features that are not characteristic for a bulk superconductor. SC may well exist within the AFM domains. The coupling of the order parameter through the FM domains can consequentially be established via the Josephson effect resulting in a phase-lock transition and "bulk" intra-grain superconductivity over the whole grain. This scenario is quit similar to that discussed for the granular ceramic samples in the previous paragraphs. However, the characteristic length scale is different (sub-grain size) and the "intra-grain granularity" is of magnetic origin, i.e. separation between AFM (superconducting) and FM (non-superconducting) domains.

To verify the physical model we focus onto the influence of a small external magnetic field on the intra-grain superconducting properties of Ru-1212 and Ru-1222. The magnetic penetration depth, $\lambda$, is an important characteristic quantity and can be deduced from ac susceptibility measurements of powdered samples. For a magnetically aligned powder of superconducting particles with diameter $d$ the real part of the ac susceptibility is expressed as [59]

$$\chi' = -3/(8\pi)\{1 - 6(\lambda/d)\coth(d/2\lambda) + 12(\lambda/d)^2\} \approx (1/500)(d/\lambda)^2 \text{ , for } d<2\lambda .$$

The formula basically presents a scaling relation expressing $\chi'$ as a function of the reduced variable $d/\lambda$. It can be used to estimate the penetration depth $\lambda$ from systematic measurements of $\chi'$ of sorted powders. In randomly oriented powders a geometric correction factor of 1/3 has to be considered to take account of the SC anisotropy. Furthermore, the particle size of sorted powders usually spreads around an average value that was estimated in our experiments by averaging over the measured size distribution, $d=[\Sigma d_i^5/\Sigma d_i^3]^{1/2}$, with $d_i$ the diameter of an individual particle, where the largest particles found still being smaller than $2d$.

The above formula was verified using a standard $YBa_2Cu_3O_{6.4}$ ceramic sample ground into powders with different average particle sizes between 0.75 and 2.1 µm [58]. The fit of $\chi'$ to the formula yields $\lambda=0.42$ µm and $T_c=42$ K. The estimated penetration depth is in good agreement with literature data on YBCO and the $T_c$ is close to the value of 40 K measured for the bulk sample.

The same procedure was applied to $RuSr_2GdCu_2O_8$. Since these materials show a magnetic background (Fig. 19) measures have been taken to estimate that background signal and to correct the $\chi'$ accordingly. Fig. 20 shows the raw data of the ac susceptibility measurement. The magnetic background was estimated based on the 0.3 µm data (It was shown that the magnetic background was independent of the particle size [58] if the different powders were prepared from one and the same starting material). The intra-grain penetration depth was estimated as $\lambda(0 K) = 3$ µm. This value is unusually large compared with other bulk high-$T_c$ materials, e.g. the YBCO sample used as a reference. However, the large $\lambda$ is in accordance with the very small intra-grain diamagnetic signal (corresponding to about 0.2 % volume fraction) observed in all bulk samples. A natural explanation can be given by employing the phase separation model discussed above. $T_c$ only represents the phase-lock temperature of a Josephson junction array (JJA) coupled through links between the AFM domains, if Ru1212 is not homogeneous on a nanoscopic scale because of the coexistence of AFM and FM domains. The intra-grain penetration depth may then be considered as an indirect measure of the (average) Josephson coupling strength. This physical picture can also explain the variations in $T_c$ depending on subtle details of the materials synthesis [8]. Since in Ru-1212 the doping level is rather fixed (c.f. discussion in Section 4.1) the variance in $T_c$ of different samples has to be related to the details of the magnetic nanostructure and the resulting JJA below $T_c$. In fact, extending the analysis described above to different $RuSr_2GdCu_2O_8$ samples with $T_c$ varying between 30 K and 40 K it was found that the penetration depth increased from 2-3 μm ($T_c$ = 40 K) to > 6 μm ($T_c$ = 30 K). This suggests a correlation between $T_c$ and $\lambda$. This correlation can be understood within the picture of intra-grain JJA's. For smaller (average) Josephson coupling $T_c$ is expected to be lower and $\lambda$ will be larger.

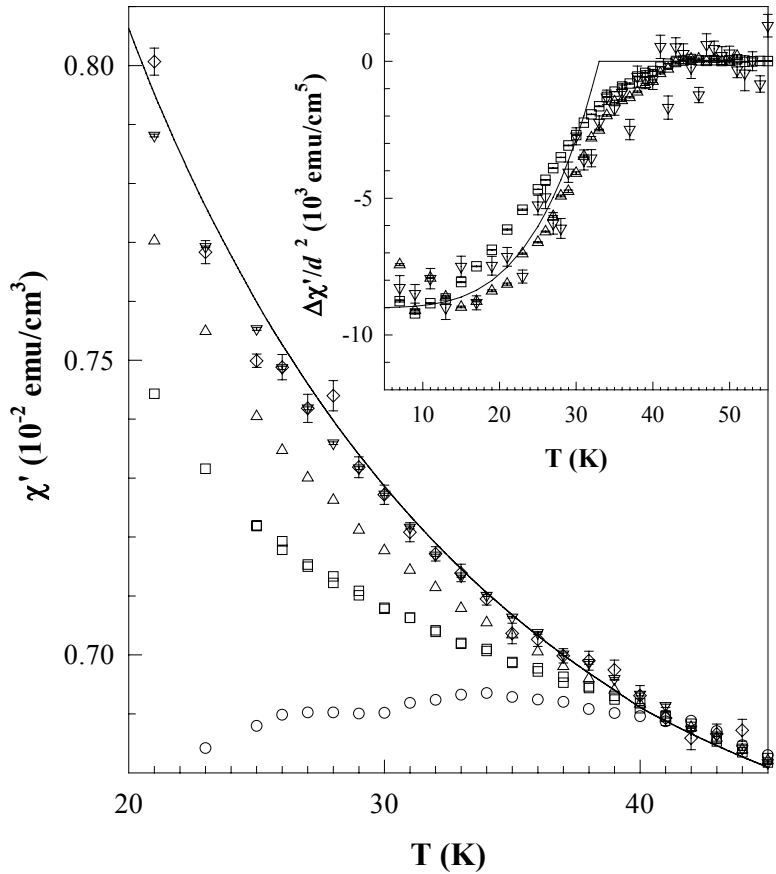

**Fig. 20**: ac susceptibility of sorted powders of $RuSr_2GdCu_2O_8$. The different symbols (top to bottom) mark data for $d$=0.3, 0.8, 1.5, 3, and 8 μm. The inset shows that the reduced $\Delta\chi'/d^2$ fulfils the scaling relation with the solid line representing the formula using $\lambda(0\ K) = 3$ μm.

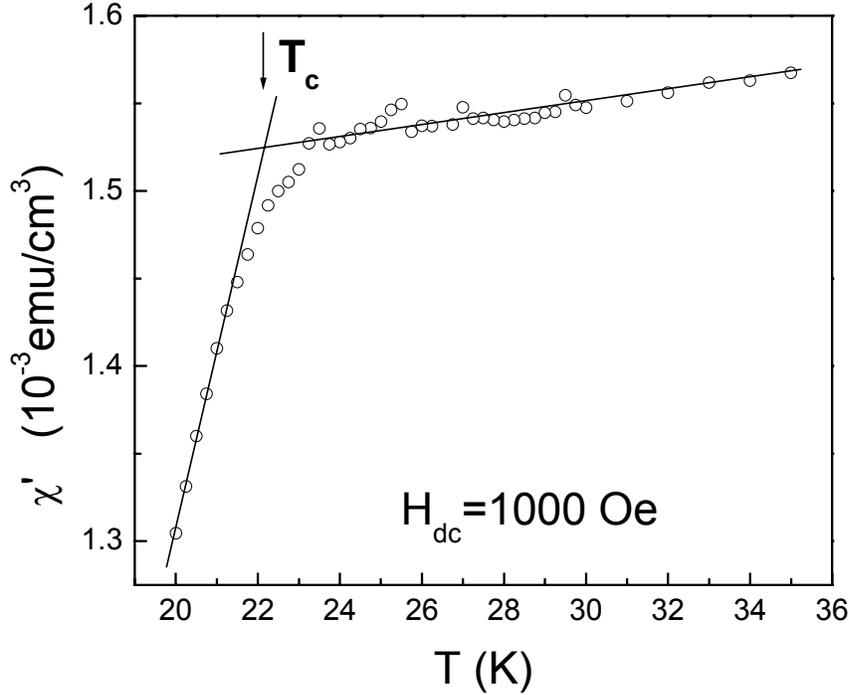

**Fig. 21**: AC susceptibility of $RuSr_2EuCu_2O_8$ in a dc bias field of 1000 Oe. The intra-grain superconducting transition temperature $T_c$ is determined from the intersection of the two straight lines.

To further support our physical picture of an intra-grain JJA we measure the magnetic field dependence of $T_c$ using ac susceptibility as well as resistivity ($\rho$) experiments. The experiments are conducted using a ceramic sample of $RuSr_2EuCu_2O_8$. The substitution of Gd by Eu reduces the paramagnetic background of Gd, and is beneficial for resolving the small intra-grain diamagnetic signal. The magnetic and superconducting transition temperatures of this compound are a few degree lower than the corresponding critical temperatures for $RuSr_2GdCu_2O_8$ but the physical properties of the FM and SC states are very similar [7]. As shown before, $T_c$ is well resolved in $\chi'$ and $\rho$. In susceptibility measurements $T_c$ is defined as the intersection point between the linear extensions of the high temperature signal and the intra-grain diamagnetic drop (shown in Fig. 21). A dc bias magnetic field up to 1 Tesla is applied. The intra-grain drop of the resistivity can be separated by a deconvolution of the two peaks observed in the derivative $d\rho/dT$. In the example of Fig. 22 the deconvolution is demonstrated by fitting $d\rho/dT$ at 500 Oe to two Gaussian shaped peaks (dotted lines). The intra-grain $T_c$ is defined by the 95 % drop of the intra-grain resistance (double arrow in Fig. 22 b).

The magnetic field dependence of $T_c$ as obtained from both measurements (Fig. 23) shows a very steep decrease at low field (about 100 K/Tesla). This strong decrease of $T_c(H)$ is not expected for a bulk, homogeneous superconductor. Ginzburg-Landau theory, for example, predicts a slower decrease for small fields and an opposite curvature over the whole temperature range (see inset of Fig. 23).

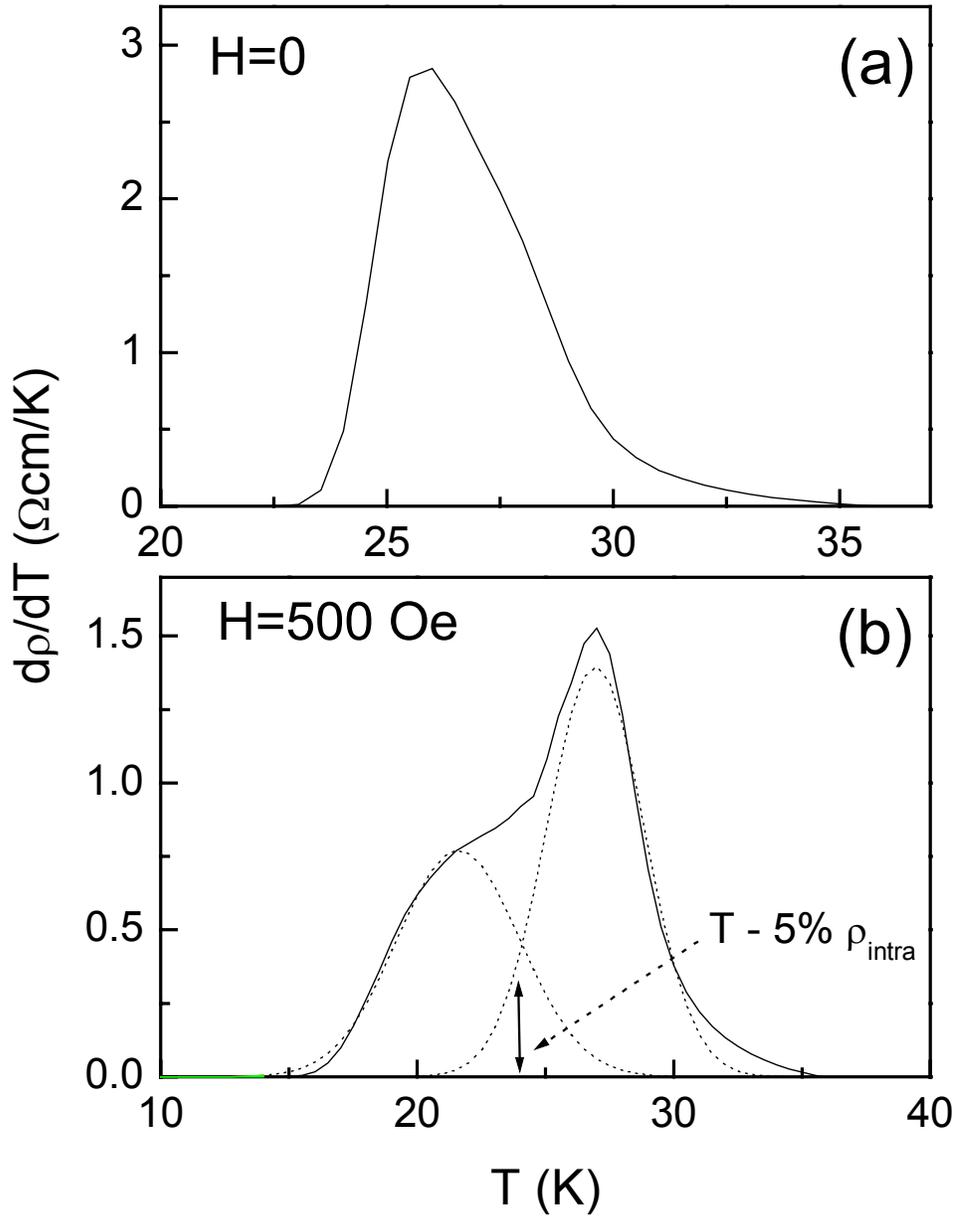

**Fig. 22**: Derivative of the resistivity at the superconducting transition. (a) Zero magnetic field. (b) H=500 Oe. Part (b) shows the superposition of the inter- and intra-grain peaks and the definition of $T_c$ at the 95 % drop of the intra-grain resistivity.

The data of Fig. 23, however, show the typical field dependence of the critical temperature of a Josephson junction array of coupled superconducting domains. This is in further support of the phase separation model. Within this physical picture the thermodynamic superconducting transition in the AFM domains is expected at a temperature higher than $T_c$. Despite extensive efforts to search for an anomaly related to this thermodynamic transition in magnetic susceptibility experiments a convincing signature of it is still missing. However, transport measurements show that the drop of resistivity and of the Seebeck coefficient sets in at a slightly higher temperature than the estimated intra-grain $T_c$. This is best documented in the derivative $d\rho/dT$ shown in Fig. 14. Below the magnetic transition $d\rho/dT$ is a perfectly linear function of T over a wide temperature range (60 to 130 K). The deviation from linearity at the low-T side might be related to the onset of superconductivity in the AFM domains (inset in Fig. 14). Further work needs to be conducted to explore this matter in more detail.

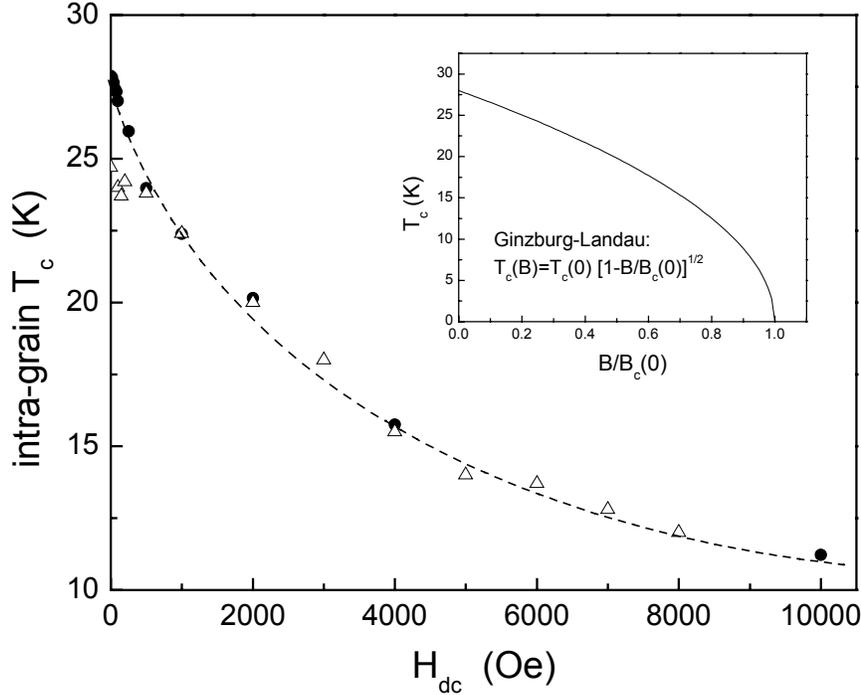

**Fig. 23**: Magnetic field dependence of the intra-grain $T_c$ of $RuSr_2EuCu_2O_8$ as determined from resistivity (open triangles) and ac susceptibility data (closed circles). The inset shows the Ginzburg-Landau dependence for a bulk superconductor.

From the foregoing discussions and from the observation that the critical temperatures for the onset of superconductivity in Ru-1212 and Ru-1222 compounds vary appreciably for different samples with the same nominal chemical composition there arises the question about the most relevant parameters determining $T_c$ and the superconducting properties in these materials. Most high-$T_c$ compounds follow the universal relation between $T_c$ and the hole density $p$ in the $CuO_2$ planes:

$$T_c = T_{c,max} [1 - 82(p-0.16)^2] .$$

$T_{c,max}$ is the maximum $T_c$ in the system at the optimal hole concentration of $p=0.16$. Since we have seen that in Ru-1212 the oxygen content and the hole density can barely be controlled by oxygen annealing we focus our attention onto the Ru-1222 structure in which doping and oxygen content can be controlled by annealing the samples in high-pressure oxygen atmosphere. Three samples from the same batch of $RuSr_2(Gd_{0.7}Ce_{0.3})_2Cu_2O_{10+\delta}$ have been prepared for susceptibility and transport measurements. The first sample (A) was taken as synthesized, the second (B) and third (C) samples were annealed for 2 hours at 600 °C at oxygen pressures of 20 bar and 300 bar, respectively. The oxygen annealing results in an increase of the intra-grain $T_c$ from 26 to 40 K (as determined from dc susceptibility as well as transport measurements) [29]. The carrier densities of these samples are calculated from the room temperature Seebeck coefficient (shown in Fig. 15) as: $p_A = 0.102$, $p_B = 0.109$, and $p_C = 0.118$. This is a very moderate increase of hole density by only 0.016 from sample A to C. To verify these values we have also estimated the change of the oxygen content with the $O_2$ annealing. Heating samples A and C to 800 °C in a gas-effusion cell the released oxygen was measured by both a pressure gauge and a mass spectrometer [60]. The estimated difference in the oxygen content between A and C of $\Delta\delta \approx 0.015$ is in excellent agreement with the difference in hole density deduced from the Seebeck coefficients.

The observed increase of $T_c$ by 14 K can barely be explained as a sole carrier density effect. Using the numbers $p_A$ and $p_C$ with a value of $T_c \approx 40$ K for sample C the expected difference of $T_c$ between samples A and C would not be larger than 6 K (we used the universal formula relating $T_c$ and $p$). Therefore, the change of carrier density induced by oxygen annealing cannot be the dominant factor determining the change of $T_c$. The enhancement of $T_c$ from samples A to C is accompanied by a systematic increase of the diamagnetic signal in FC as well as ZFC dc susceptibility data [29]. For example, the drop of the magnetic moment, $\Delta M_{ZFC}$, measured at 5 Oe increases from 0.15 (sample A) to 0.2 (sample B) and to 0.3 emu/cm$^3$ (sample C). This implies a change of the magnetic penetration depth, i.e. the increase of $T_c$ and $\Delta M_{ZFC}$ is related to the increase of $1/\lambda^2$ (the superfluid density of a bulk superconductor).

To verify the conclusion we measure the intra-grain penetration depth according to the same procedure applied to $RuSr_2GdCu_2O_8$ (see above). The ac susceptibility of sorted powders of $RuSr_2(Gd_{0.7}Ce_{0.3})_2Cu_2O_{10+\delta}$ was measured and $\lambda$ was calculated from the particle size dependence using a regression procedure that includes geometrical corrections and a magnetic background contribution to $\chi'$ (see [29] for details). With no arbitrarily adjustable parameter the regression allows to determine $\lambda(T)$ within an estimated error of less than 30 %. Fig. 24 shows the results obtained for samples A and C. The zero temperature $1/\lambda^2$ of the as synthesized sample A ($\approx 3.5$ μm$^{-2}$) is about twice the value of that of sample C ($\approx 7$ μm$^{-2}$) indicating that the oxygen annealing effect on $1/\lambda^2$ is more pronounced than that on the carrier density $p$, i.e. intra-grain Josephson junctions play the dominating role. This view is further supported by the fast decrease of $T_c$ if a small dc magnetic field is applied. The initial slope of $T_c(H)$ is of the same order (about 100 K/Tesla) as discussed above for Ru-1212 (Fig. 23).

We have treated a large number of Ru-1222 samples under different annealing conditions including pure argon and mixed argon + oxygen atmospheres and applied the same analysis to extract the $1/\lambda^2$ and $T_c$. The range of $1/\lambda^2$ extends from 0.3 to about 7 μm$^{-2}$ with a $T_c$ stretching from 15 K to 40 K. The large increase of $1/\lambda^2$ by a factor of more than 20 is significantly higher than expected from the change of both, $p$ and $T_c$, and it deserves further attention. In principle, $1/\lambda^2$ may be affected by the normal state effective mass $m^*$ of the carriers, scattering resulting in pair-breaking of the Cooper pairs, and the intra-grain granularity discussed above (in the latter case $1/\lambda^2$ is determined by the Josephson coupling between the SC domains that can be sensitive to the annealing conditions).

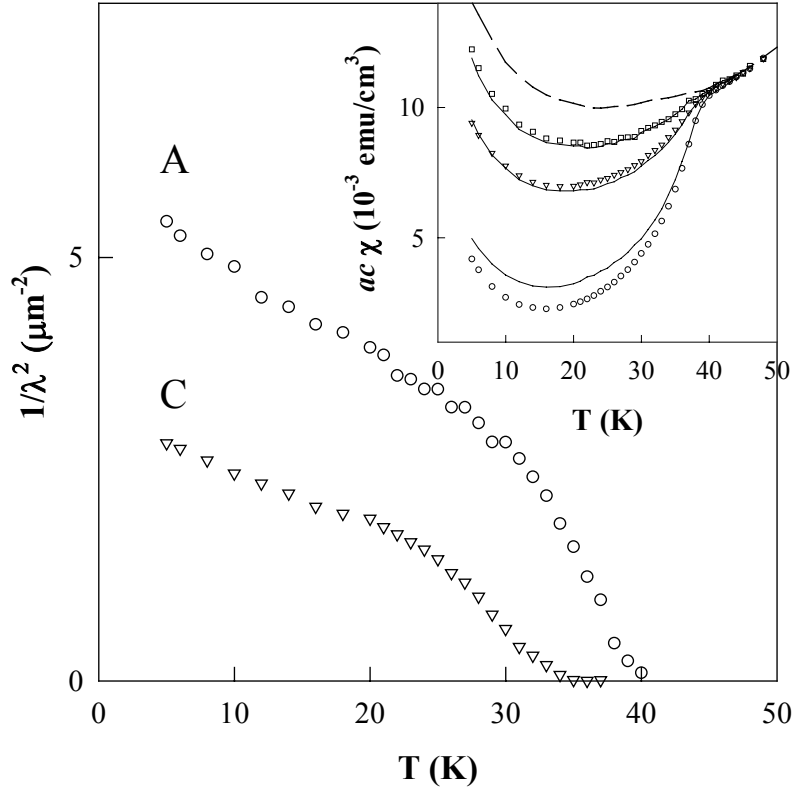

**Fig. 24**: The intra-grain $1/\lambda^2$ for $RuSr_2(Gd_{0.7}Ce_{0.3})_2Cu_2O_{10+\delta}$ before (A) and after (C) 300 bar oxygen annealing. The inset shows the measured $\chi'$ for powders made from sample C with particle size of 1.9 μm (o), 1.3 μm ($\nabla$), and 0.9 μm ($\square$). The solid lines are the fits to the data and the dashed line indicates the estimated magnetic background signal.

The effective mass effect can be excluded since the room temperature resistivity of samples A and C shows only a minor difference. For an underdoped bulk high-$T_c$ superconductor Uemura et al. [61] proposed that the critical temperature $T_c$ is proportional to the superfluid density. This relation (Uemura-line) appears to be universal and was also shown to hold for Zn substituted high-$T_c$ superconductors where $T_c$ decreased due to pair-breaking effects [62]. In Fig. 25 we plot $1/\lambda^2$ versus $T_c$. Obviously, the data for our Ru-1222 samples do not fit to the Uemura-line but lay all far left/above the proposed universal line. The data are comparable with our results for Ru-1212 where samples with penetration depths as large as 3 μm still have a $T_c > 20$ K [58]. This provides strong evidence that the differences in $1/\lambda^2$ are not generated by pair-breaking mechanisms. Although $T_c$ still increases linearly with $1/\lambda^2$ the distinct offset from the Uemura-line provides compelling evidence that the intra-grain superconducting state cannot be considered as "homogeneous". This is in favor of our suggestion that the SC state within a grain is more likely described by a JJA of coupled SC sub-grain domains. The $T_c$ of a JJA is proportional to the coupling energy of a junction. The $\lambda$ of the junction may depend on the junction length but the $T_c$ does not. Therefore, a comparably large phase-lock $T_c$ can coexist even with a long $\lambda$ (or a small value of $1/\lambda^2$).

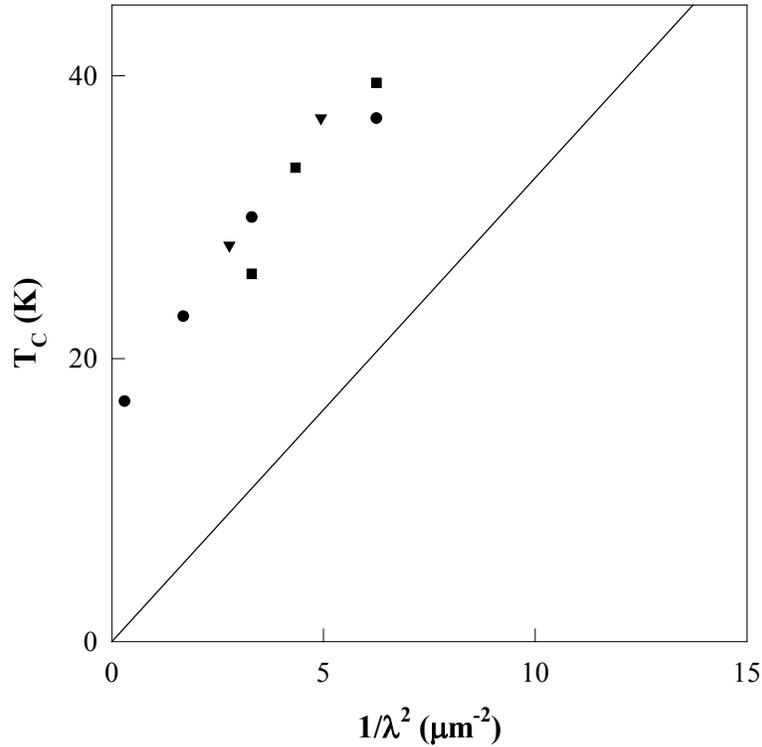

**Fig. 25**: $T_c$ vs $1/\lambda^2$ at 5 K for several annealed samples of $RuSr_2(Gd_{0.7}Ce_{0.3})_2Cu_2O_{10+\delta}$. Solid line: Uemura-line.

### 4.5 Competition Between Superconducting and Ferromagnetic States

One of the big puzzles in the superconducting ferromagnets is the coexistence of both antagonistic states over a wide temperature range. Therefore, the question arises if FM and SC compete with one another or if there is no mutual interference. Based on muon spin rotation experiments it was suggested that the magnetic moments are not affected by the onset of superconductivity in Ru-1212 below $T_c \approx 45$ K [24]. However, experiments on chemical substitution (doping) of Ru-1212 show that $T_c$ and $T_m$ are changed in an opposite way. Decreasing the hole density by partially replacing Gd with Ce [63] or Sr with La [64] results in a decrease of $T_c$ (with an ultimate suppression of superconductivity) and an increase of the FM transition temperature. This observation is in line with the fact that the magnetic transition temperature for non-superconducting Ru-1212 is a few degree higher than that of superconducting samples [53, 65]. These results may be an indication of competition between FM and SC states. However, the chemical substitution affects several parameters simultaneously. Besides the change of carrier density it may also introduce disorder, reduce the magnetic coupling in the $RuO_2$ layers, and cause changes of the microstructure.

The use of thermodynamic variables to tune the FM and SC states appears to be interesting because it usually does not change the chemistry of the compounds. We, therefore, decided to use hydrostatic pressure to investigate its effect on the intra-grain superconductivity and on the ferromagnetic state of $RuSr_2GdCu_2O_8$. As shown above, both transition temperatures ($T_c$ and $T_m$) can well be determined from resistivity as well as ac susceptibility measurements (see Figs. 14, 17, 18, 22). The two transitions were investigated at pressures up to 2 GPa by measuring simultaneously the real part of the ac susceptibility and the resistivity. A dual coil system was mounted to a ceramic pellet of $RuSr_2GdCu_2O_8$ and four platinum wires were attached to the sample for resistance measurements. Both, $\chi'(T)$ and $\rho(T)$, were measured in parallel employing the resistance and mutual inductance bridge LR 700 (Linear Research). Hydrostatic pressure was generated in a beryllium-copper piston-cylinder clamp. The sample was mounted in a Teflon container filled with a 1:1 mixture of Fluorinert FC70 and FC77 as a pressure-transmitting medium. The pressure was measured *in situ* at about 7 K by monitoring the shift of the superconducting transition temperature of a small piece of high purity (99.9999 %) lead. The temperature was measured by a thermocouple inside the Teflon pressure cell as well as a germanium resistor built into the BeCu clamp close to the sample position.

It is particularly important in this experiment that we can separate the inter- and intra-grain SC transitions in order to discuss the "intrinsic" pressure effect. There is an additional contribution to the pressure coefficient of the inter-grain critical temperature ($T_p$) due to the fact that pressure naturally compresses the polycrystalline sample and, therefore, improves the grain-grain connectivity resulting in an increase of the inter-grain Josephson coupling. This effect is not "intrinsic" and should not interfere with the intra-grain shift of $T_c$. Since we have shown that $T_p$ and $T_c$ are well distinguished in our experiments the application of pressure can be used to reveal any correlations between the intra-grain superconductivity and magnetism.

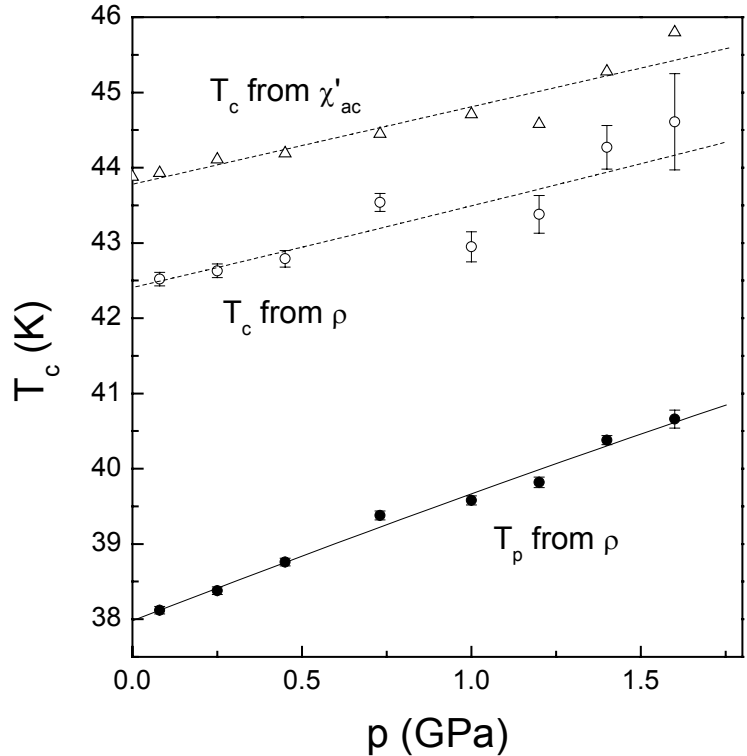

**Fig. 26**: Pressure dependence of $T_c$ estimated from $\chi'$ ($\Delta$) and $\rho$ (o) and $T_p$ (•).

The pressure dependence of $T_c$ and $T_p$ is shown in Fig. 26. The pressure coefficients of $T_c$ of 1.02 K/GPa and 1.06 K/GPa as obtained from $\chi'$ and $\rho$, respectively, are in excellent agreement. The small offset between the two data sets is due to the different criteria to estimate $T_c$ from $\chi'$ and $\rho$. The larger pressure shift of the inter-grain phase-lock temperature, $dT_p/dP = 1.8$ K/GPa reflects the abovementioned pressure-induced improvement of the grain-grain contacts. The pressure shift of the magnetic $T_m$ (Fig. 27), $dT_m/dP = 6.7$ K/GPa, is distinctly larger than that of $T_c$. Comparing the relative enhancements, $d\ln T_c/dP = 0.025$ GPa$^{-1}$ and $d\ln T_m/dP = 0.054$ GPa$^{-1}$, it becomes obvious that the magnetic $T_m$ increases about twice as fast with P as the superconducting $T_c$.

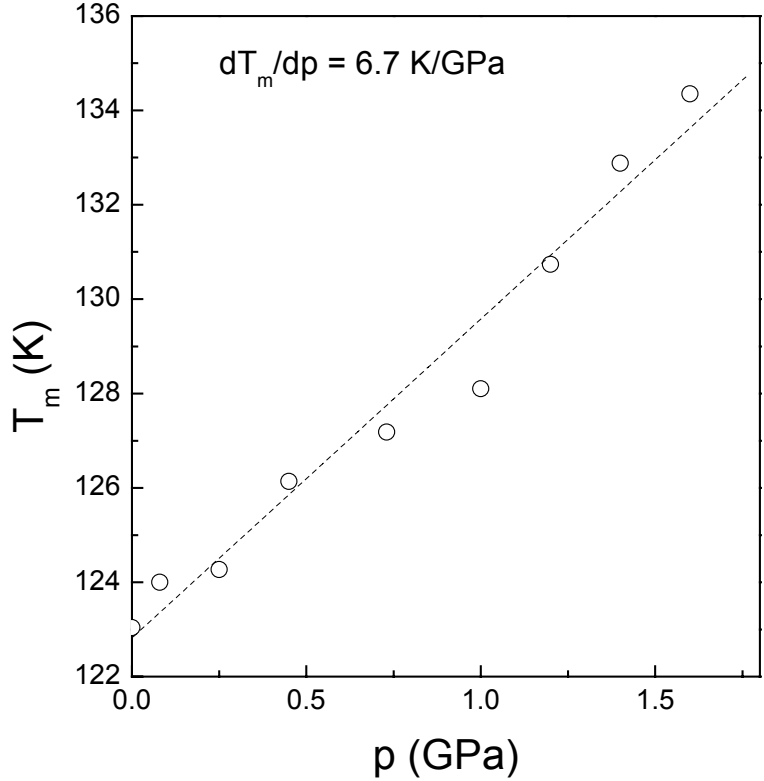

**Fig. 27**: Pressure dependence of the ferromagnetic transition temperature, $T_m$.

For a further exploration of the pressure effect in Ru1212 we compare the pressure coefficient of $T_c$ with that of other high-$T_c$ compounds in a similar doping state. Ru-1212 is a typical underdoped material with a hole concentration of about 0.07 (see discussion in Section 4.1). The pressure coefficients of other underdoped cuprates were recently estimated for a number of different compounds, e.g. La$_{2-x}$(Sr,Ba)$_x$CuO$_4$ [66], La$_2$CuO$_{4+\delta}$ [67], YBa$_2$Cu$_3$O$_{7-\delta}$ [68], or YBa$_2$Cu$_{3-x}$M$_x$O$_{7-\delta}$ [69], and typical values ranging from 3 to 4 K/GPa or even higher were reported. This number can be considered as a universal value for underdoped cuprates. The equivalent value for Ru-1212 of only 1 K/GPa is at least 3 to 4 times lower than the pressure coefficient expected in the particular doping state. This unusually small $dT_c/dP$ of RuSr$_2$GdCu$_2$O$_8$ leads us to the conclusion that it is primarily an effect of the enhancement of the FM order by pressure. It is then an immediate consequence of an apparent competition of the FM and SC phases. Due to this competition the stronger enhancement of the magnetic phase results in a reduced (as compared to other high-$T_c$ compounds) pressure effect on $T_c$. This effect can be understood within our physical picture of intra-grain FM and AFM phase separation. The P-induced increase of $T_m$ indicates an enhanced ferromagnetic exchange interaction in the non-superconducting FM domains. This will result in a reduced Josephson coupling between the superconducting (AFM) domains and a relative reduction of the intra-grain $T_c$. The intrinsic pressure effect on the intra-grain superconductivity is, therefore, reduced to the small value of 1 K/GPa observed in our experiments. Our high-pressure experiments provide additional indirect support for the explanation of the intra-grain superconductivity as the phase-lock transition of a JJA that originates from the magnetic phase separation into AFM and FM domains.

## 5. SUMMARY

The magnetic and superconducting properties of a class of newly discovered compounds, superconducting ferromagnets, have been extensively investigated. Magnetic susceptibility measurements (dc and ac) as well as electrical transport experiments (resistivity and thermoelectric power) were employed to understand the unusual magnetic and superconducting properties and, in particular, the coexistence and competition between these antagonistic states of matter.

We conclude that in both ruthenium based high-$T_c$ structures, Ru-1212 and Ru-1222, the magnetic order is very complex and shows the typical signatures of phase separation into anti-ferromagnetic and ferromagnetic domains on a sub-grain scale. Accordingly, the intra-grain superconductivity in these compounds is understood as a phase-lock transition between superconducting and anti-ferromagnetic domains with a weak Josephson like coupling across the ferromagnetic (but non-superconducting) domains. The exotic and yet not well understood properties of the superconducting state can be explained by the model. For example, the extremely small diamagnetic signal, the small and sometimes missing Meissner signal, the large intra-grain penetration depth, the unusual magnetic field dependence of the intra-grain $T_c$, and the pressure effects on the magnetic and superconducting states can be explained within the physical picture described above and lend strong support to the model of magnetic phase separations appearing in superconducting ferromagnets.

**Acknowledgements**

This work is supported in part by NSF Grant No. DMR-9804325, the T.L.L. Temple Foundation, the John J. and Rebecca Moores Endowment, and the State of Texas through the Texas Center for Superconductivity at the University of Houston and at Lawrence Berkeley Laboratory by the Director, Office of Energy Research, Office of Basic Energy Sciences, Division of Materials Sciences of the U.S. Department of Energy under Contract No. DE-AC03-76SF00098.

REFERENCES

[1] L. Bauernfeind, W. Widder, and H. F. Braun, Physica C **254** [1995] 151.
[2] A. Ono, Jpn. J. Appl. Phys. **34** [1995] L 1121.
[3] I. Felner, U. Asaf, Y. Levi, and O. Millo, Phys. Rev. B **55** [1997] R3374.
[4] I. Felner, U. Asaf, S. Reich, and Y. Tsabba, Physica C **311** [1999] 163.
[5] K. Otzschi, T. Mizukami, T. Hinouchi, J. Shimoyama, and K. Kishio, J. Low Temp. Phys. **117** [1999] 855.
[6] C. W. Chu, Y. Y. Xue, Y. S. Wang, A. K. Heilman, B. Lorenz, R. L. Meng, J. Cmaidalka, L. M. Dezaneti, J. Supercond. **13** [2000] 679.
[7] R. L. Meng, B. Lorenz, Y. S. Wang, J. Cmaidalka, Y. Y. Xue, and C. W. Chu, Physica C **353** [2001] 195.
[8] B. Lorenz, R. L. Meng, J. Cmaidalka, Y. S. Wang, J. Lenzi, Y. Y. Xue, and C. W. Chu, Physica C **363** [2001] 251.
[9] C. Bernhard, J. L. Tallon, E. Brücher, and R. K. Kremer, Phys. Rev. B **61** [2000] R14960.
[10] B. T. Matthias, H. Suhl, and E. Corenzwit, Phys. Rev. Lett. **1** [1958] 92.
[11] Superconductivity in Ternary Compounds II, edited by M. B. Maple and O. Fisher (Springer-Verlag, Berlin, 1982).
[12] H. Eisake, H. Takagi, R. J. Cava, B. Batlogg, J. J. Krajewski, W. F. Peck, Jr., K. Mizuhashi, J. O. Lee, and S. Uchida, Phys. Rev. B **50** [1994] 647.
[13] L. N. Bulaevskii, A. I. Buzdin, M. L. Kulic, and S. V. Panjukov, Adv. Phys. **34** [1985] 175.
[14] S. K. Sinha, H. A. Mook, D. G. Hinks, and G. W. Crabtree, Phys. Rev. Lett. **48** [1982] 950.
[15] W. Thomlinson, G. Shirane, J. W. Lynn, and D. E. Moncton, Superconductivity in ternary compounds II, edited by M. B. Maple and O. Fisher (Springer-Verlag, Berlin, 1982), p. 99.
[16] J. W. Lynn, J. A. Gotaas, R. W. Erwin, R. A. Ferrel, J. K. Bhattacharjee, R. N. Shelton, and P. Klevins, Phys. Rev. Lett. **52** [1984] 133.
[17] W. E. Pickett, R. Weht, and A. B. Shick, Phys. Rev. Lett. **83** [1999] 3713.
[18] P. Fulde and R. A. Ferrell, Phys. Rev. **135** [1964] A550.
A. I. Larkin and Yu. N. Ovchinnikov, Sov. Phys. JETP **20** [1965] 762.
[19] R. A. Ferrel, J. K. Bhattacharjee, and A. Bagchi, Phys. Rev. Lett. **43** [1997] 154.
[20] H. Matsumota, H. Umezawa, and M. Tachiki, Solid State Commun. **31** [1979] 157.
[21] H. S. Greenside, E. I. Blount, and C. M. Varma, Phys. Rev. Lett. **46** [1981] 49.
[22] E. B. Sonin and I. Felner, Phys. Rev. B **57** [1998] R14000.
[23] C. W. Chu, Y. Y. Xue, S. Tsui, J. Cmaidalka, A. K. Heilman, B. Lorenz, and R. L. Meng, Physica C **335** [2000] 231.
[24] C. Bernhard, J. L. Tallon, Ch. Niedermayer, Th. Blasius, A. Golnik, E. Brücher, R. K. Kremer, D. R. Noakes, C. E. Stronach, and E. J. Ansaldo, Phys. Rev. B **59** [1999] 14099.
[25] J. W. Lynn, B. Keimer, C. Ulrich, C. Bernhard, and J. L. Tallon, Phys. Rev. B **61** [2000] R14964.
[26] O. Chmaissen, J. D. Jorgensen, H. Shaked, P. Dollar, and J. L. Tallon, Phys. Rev. B **61** [2000] 6401.
[27] J. D. Jorgensen, O. Chmaissen, H. Shaked, S. Short, P. W. Klamut, B. Dabrowski, and J. L. Tallon, Phys. Rev. B **63** [2001] 054440.

[28] Y. Tokunaga, H. Kotegawa, K. Ishida, Y. Kitaoka, H. Takagiwa, and J. Akimitsu, Phys. Rev. Lett. **86** [2001] 5767.
[29] Y. Y. Xue, B. Lorenz, A. Baikalov, D. H. Cao, Z. G. Li, and C. W. Chu, Phys. Rev. B **66** [2002] 014503.
[30] Y. Y. Xue, R. L. Meng, J. Cmaidalka, B. Lorenz, L. M. Dezaneti, A. K. Heilman, and C. W. Chu, Physica C **341-348** [2000] 459.
[31] M. Pozek, A. Dulcic, D. Paar, A. Hamzic, M. Basletic, E. Tafra, G. V. M. Williams, and S. Krämer, Phys. Rev. B **65** [2002] 17514.
[32] I. Felner, U. Asaf, Y. Levi, and O. Millo, Physica C **334** [2000] 141.
[33] I. Felner, U. Asaf, S. D. Goren, and C. Korn, Phys. Rev. B **57** [1998] 550.
[34] J. L. Tallon, J. W. Loram, G. V. M. Williams, and C. Bernhard, Phys. Rev. B **61** [2000] R6471.
[35] X. H. Chen, Z. Sun, K. Q. Wang, S. Y. Li, Y. M. Xiong, M. Yu, and L. Z. Cao, Phys. Rev. B **63** [2001] 064506.
[36] A. C. McLaughlin, W. Zhou, J. P. Attfield, A. N. Fitch, and J. L. Tallon, Phys. Rev. B **60** [1999] 7512.
[37] G. M. Kuz'micheva, V. V. Luparev, E. P. Khlybov, I. E. Kostyleva, A. S. Andeenko, and K. N. Gavrilov, Physica C **350** [2001] 105.
[38] G. V. M. Williams, A. C. McLaughlin, J. P. Attfield, S. Krämer, and H. K. Lee, cond-mat/0108521, Aug. 30, 2001, unpublished.
[39] H. Takigawa, J. Akimitsu, H. Kawano-Furukawa, and H. Yoshizawa, J. Phys. Soc. Jpn. **70** [2001] 333.
[40] K. I. Kumagai, S. Takada, and Y. Furukawa, Phys. Rev. B **63** [2001] 180509.
[41] A. Butera, A. Fainstein, E. Winkler, and J. Tallon, Phys. Rev. B **63** [2001] 054442.
[42] Y. Y. Xue, D. H. Cao, B. Lorenz, and C. W. Chu, Phys. Rev. B **65** [2002] R020511.
[43] A. Fainstein, E. Winkler, A. Butera, and J. Tallon, Phys. Rev. B **60** [2000] R12597.
[44] Cz. Kapusta, P. C. Riedi, M. Sikora, and M. R. Ibarra, Phys. Rev. Lett. **84** [2000] 4216.
[45] A. P. Ramirez, P. Schiffer, S.-W. Cheong, C. H. Chen, W. Bao, T. T. M. Palstra, P. L. Gammel, D. J. Bishop, and B. Zegarski, Phys. Rev. Lett. **76** [1996] 3188.
[46] M. R. Ibarra, J. M. De Teresa, J. Blasco, P. A. Algarabel, C. Marquina, J. Garcia, J. Stankiewicz, and C. Ritter, Phys. Rev. B **56** [1997] 8252.
[47] A similar maximum of the high field differential susceptibility has been reported in Ru1212Eu. However, its temperature of 145 K is only slightly higher than the corresponding low-H FM-like transition temperature of 138 K, and was assigned to the same canted AFM transition in Ref. [41].
[48] I. Felner, U. Asaf, and E. Galstyan, cond-mat/0111217, Nov. 12, 2001, unpublished.
[49] Y. Y. Xue, D. H. Cao, B. Lorenz, and C. W. Chu, unpublished.
[50] For example, P. Allia, M. Coisson, P. Tiberto, F. Vinai, M. Knobel, M. A. Novak, and W. C. Nunes, Phys. Rev. B **64** [2001] 144420.
[51] O. Chauvert, G. Goglio, P. Molinie, B. Corraze, and L. Brohan, Phys. Rev. Lett. **81** [1998] 1102.
[52] S. D. Obertelli, J. R. Cooper, and J. L. Tallon, Phys. Rev. B **46** [1992] 14928.
[53] P. W. Klamut, B. Dabrowski, S. Kolesnik, M. Maxwell, and J. Mais, Phys. Rev. B **63** [2001] 224512.
[54] C. T. Lin, B. Liang, C. Ulrich, and C. Bernhard, Physica C **364-365** [2001] 373.
[55] B. Lorenz, Y. Y. Xue, R. L. Meng, and C. W. Chu, Phys. Rev. B **65** [2002] 174503.

[56] I. Felner, U. Asaf, F. Ritter, P. W. Klamut, and B. Dabrowski, Physica C **364-365** [2001] 368.
[57] A. Fainstein, P. Etchegoin, H. J. Trodahl, and J. L. Tallon, Phys. Rev. B **61** [2001] 15468.
[58] Y. Y. Xue, B. Lorenz, R. L. Meng, A. Baikalov, and C. W. Chu, Physica C **364-365** [2001] 251.
[59] A. Porch, J. R. Cooper, D. N. Zheng, J. R. Waldram, A. M. Campbell, and P. A. Freeman, Physica C **214** [1993] 350.
[60] A. Hamed, R. Ortiz, H. H. Feng, Z. G. Li, P. H. Hor, Y. Y. Xue, Y. Y. Sun, Q. Xiong, Y. Cao, and C. W. Chu, Phys. Rev. B **54** [1996] 682.
[61] Y. J. Uemura, G. M. Luke, B. J. Sternlieb, J. H. Brewer, J. F. Carolan, W. N. Hardy, R. Kadono, J. R. Kempton, R. F. Kiefl, S. R. Kreitzman, P. Mulhern, T. M. Riseman, D. L. Williams, B. X. Yang, S. Uchida, H. Takagi, J. Gopalakrishnan, A. W. Sleight, M. A. Subramanian, C. L. Chien, M. Z. Cieplak, G. Xiao, V. Y. Lee, B. W. Statt, C. E. Stronach, W. J. Kossler, and X. H. Yu, Phys. Rev. Lett. **62** [1989] 2317.
[62] B. Nachumi, A. Keren, K. Kojima, M. Larkin, G. M. Luke, J. Merrin, O. Tchernyshöv, Y. J. Uemura, N. Ichikawa, M. Goto, and S. Uchida, Phys. Rev. Lett. **77** [1996] 5421.
[63] P. W. Klamut, B. Dabrowski, J. Mais, and M. Maxwell, Physica C **350** [2001] 24.
[64] P. Mandal, A. Hassen, J. Hemberger, A. Krimmel, and A. Loidl, Phys. Rev. B **65** [2002] 144506.
[65] P. W. Klamut, B. Dabrowski, S. M. Mini, M. Maxwell, S. Kolesnik, J. Mais, A. Shengelaya, R. Khasanov, I. Savic, H. Keller, T. Graber, J. Gebhardt, P. J. Viccaro, and Y. Xiao, Physica C **364-365** [2001] 313.
[66] Q. Xiong, Ph.D. Thesis, University of Houston, 1993.
[67] B. Lorenz, Z. G. Li, T. Honma, and P.-H. Hor, Phys. Rev. B **65** [2002] 144522.
[68] W. H. Fietz, R. Quenzel, H. A. Ludwig, K. Grube, S. I. Schlachter, F. W. Hornung, T. Wolf, A. Erb, M. Kläser, and G. Müller-Vogt, Physica C **270** [1996] 258.
[69] Q. Xiong, Y. Y. Xue, J. W. Chu, Y. Y. Sun, Y. Q. Wang, P. H. Hor, and . W. Chu, Phys. Rev. B **47** [1993] 11337.